\def \SAIT #1 #2 {{\em Mem.\ Soc.\ Astron.\ It.\/} {\bf #1}, #2}
\def \MESS #1 #2 {{\em The Messenger\/} {#1}, #2}
\def \ASTRNACH #1 #2 {{ Astron. Nach.\/} { #1}, #2}
\def \AAP #1 #2 {{ A{\rm \&}A\/} {#1}, #2}
\def \AAL #1 #2 {{ A{\rm \&}A\/} {#1}, L#2}
\def \AAR #1 #2 {{ A{\rm \&}AR\/} {#1}, #2}
\def \AAS #1 #2 {{ A{\rm \&}AS\/} {#1}, #2}
\def \AJ #1 #2 {{ AJ\/} {#1}, #2}
\def \ANNREV #1 #2 {{ ARA{\rm \&}A\/} {#1},#2}
\def \APJ #1 #2 {{ ApJ\/} {#1}, #2}
\def \APJL #1 #2 {{ ApJ\/} {#1}, L#2}
\def \APJS #1 #2 {{ ApJS\/} {#1}, #2}
\def \APSS #1 #2 {{ Ap{\rm \&}SS\/} {#1}, #2}
\def \ASR #1 #2 {{ Adv. Space Res.\/} {#1}, #2}
\def \BAIC #1 #2 {{ Bull. Astron. Inst. Czechosl.\/} { #1}, #2}
\def \JSQRT #1 #2 {{ J. Quant. Spectrosc. Radiat. Transfer\/} {
#1}, #2}
\def \MN #1 #2 {{ MNRAS\/} { #1}, #2}
\def \MEM #1 #2 {{ Mem. R. Astr. Soc.\/} { #1}, #2}
\def \PLR #1 #2 {{ Phys. Lett. Rev.\/} { #1}, #2}
\def \PASJ #1 #2 {{ Publ. Astron. Soc. Japan\/} { #1}, #2}
\def \PASP #1 #2 {{ Publ. Astr. Soc. Pacific\/} { #1}, #2}
\def \NAT #1 #2 {{ Nat\/} { #1}, #2}
\def \ACTA #1 #2 {{ Acta Astron.\/} { #1}, #2}
\documentclass[numberedappendix,appendixfloats,linenumbers]{emulateapj}

\DeclareRobustCommand{\ion}[2]{%
\relax\ifmmode
\ifx\testbx\f@series
{\mathbf{#1\,\mathsc{#2}}}\else
{\mathrm{#1\,\mathsc{#2}}}\fi
\else\textup{#1\,{\mdseries\textsc{#2}}}%
\fi}

\slugcomment{To appear in ApJ.}

\shorttitle{Insights into the evolution of five isolated galaxies}
\shortauthors{Mazzei et al.}

\usepackage{graphicx, lineno}

\def\smallskip{\vskip 6pt}

\def\M12{${\rm M_{12}}$}

\begin{document}
\title{Insights into the evolution of five isolated galaxies}
\author{P. Mazzei\altaffilmark{1},
R. Rampazzo\altaffilmark{2}
A. Marino\altaffilmark{1},
G. Trinchieri\altaffilmark{3}, 
M. Uslenghi\altaffilmark{4},
A. Wolter\altaffilmark{3},
}
\altaffiltext{1}{INAF Osservatorio Astronomico di Padova, Vicolo
dell'Osservatorio 5, 35122 Padova, Italy}

\altaffiltext{2}{Universit\`a degli Studi di Padova, Dipartimento di Fisica e Astronomia 
``G. Galilei" Vicolo dell'Osservatorio 3, 35122, Padova, Italy}

\altaffiltext{3}{INAF Osservatorio Astronomico di Brera, Via Brera 26, 20121, Milano, Italy}

\altaffiltext{4}{INAF, IASF, Via Curti 12, Milano, Italy}

\date{Received / Accepted }

\begin{abstract}

 {The galaxy evolution is believed to be conditioned
by the environment. Isolated galaxies or galaxies in poor groups are an
excellent laboratory to study evolutionary mechanisms  where
effects of the environment are minimal. We present new {\it
Swift}-{\tt UVOT} data  
 in six filters, three in the ultraviolet (UV),  of five isolated galaxies aiming at shedding
light into their evolution. For all of our targets we present new UV
integrated fluxes and for some of them also new UBV magnitudes. Our
observations allow us  to improve their multi-wavelength spectral energy
distributions  extending it over about 3 orders of
magnitude in wavelength. We exploit our
smooth-particle hydro-dynamical  simulations with chemo-photometric
implementation anchored, a posteriori, to the global
multi-wavelength properties of our targets, to give insight into their
evolution. Then we compare  their evolutionary properties with those previously
derived   for several galaxies in groups.  The evolution of our targets is driven by  a  merger occurred 
several Gyrs ago, in the redshift range $0.5\leq z \leq 4.5$, not unlike   what we  have already  found for
galaxies in groups.
The merger shapes the potential well  where the gas is
accreting driving the star formation rate and the galaxy evolution.   Isolated galaxies  should not have suffered from interactions for at least 3\,Gyr. 
However, the initial merger  is still
leaving its signatures on the  properties of our targets.  Several rejuvenation episodes, triggered by {\it in situ}  accretion, are highlighted. Moreover, jelly-fish morphologies appear
as these galaxies achieve their maximum star formation rate, before their quenching phase.
}
\end{abstract}

\medskip

\section{Introduction}
\label{intro}
The research on isolated galaxies started in 1970's from the
investigation of the Palomar Observatory Sky Survey (POSS) plates
\citep[see e.g.][for a historical persepective]{Rampazzo2016}.
Conceptually, a galaxy is considered isolated if it has not experienced
any gravitational influence from neighbours, at least over the last few
billion years. In practice, isolated galaxies are sought among those
that, at first instance, do not have (obvious) nearby companions.
\citet{Karachentseva1973} compiled the first catalog of such objects,
{\it The Catalog of Isolated Galaxies} (CIG hereafter), from the  visual
examination of POSS plates adopting 2D isolation criteria, since in most
cases redshifts were not available. CIG includes 1050 candidates of all
morphological classes, although spiral candidates are predominant. Since
then,  with the increase in the number of candidates with known redshift, several
other catalogs have followed over the years, either revisions of the CIG
 \citep{Verdes2005, Verley2007a,Argudo2013} or new 3D compilations
 \citep{Argudo2015}.
\begin{table}
\centering
\scriptsize
\caption{Commonly used acronyms}
\begin{tabular}{lccccc}
\hline\hline
Acronym &  \\
\hline
BP & Brightest Point \\
BC & Blue Cloud \\
 CIG & Catalog of Isolated Galaxies \\
 CMD & Color Magnitude Diagram (NUV-r vs M$_r$) \\
 CPI & Chemo-Photometric Implementation\\
 DM & Dark Matter \\
 ETG  &  Early Type (elliptical+S0) Galaxy  \\
 FIR & Far-Infrared \\
 FUV & Far-Ultra Violet \\
 GV & Green Valley \\
 GALEX & GALaxy evolution EXplorer \\
 iETG & isolated ETG\\
 LDE & Low Density Environment \\
 LTG    &  Late Type Galaxy   \\
 RS & Red Sequence \\
 SED & Spectral Eenergy Distribution \\
 SF & Star Formation \\
 SFR &  Star Formation Rate\\
 SSFR & Specific  Star Formation Rate\\
 SPH   &  Smoothed Particle Hydrodynamics         \\
 UV  & UltraViolet \\
\hline
\end{tabular}
\label{table0}
\end{table}

During the past 30 years it has become clear that galaxy properties
e.g. morphology, star formation (SF), nuclear activity, and evolution may be
driven as strongly by initial conditions   as by environment,   that is
the subsequent dynamical processes they experienced \citep[see e.g.][and
references therein]{Boselli2006, Cappellari2016}. In this context, the
value of galaxies in isolation seems twofold. From one side, isolation
offers the possibility either to reduce or even to exclude the influence
of the environment from the study of galaxy evolution going directly to
basic ingredients. From the other, a well selected sample of isolated
galaxies would represent the baseline with which to compare the
properties of galaxies inhabiting areas of different richness, from 
which to give back insight about  assembling mechanisms at work
\citep{Boselli2006}.
There  are two
kinds of  environmental influences a) one-by-one and b) local number
 density of galaxies. A  single, sometimes difficult to identify, neighbour
can be able of a larger effect than an enhanced local  number
density of galaxies. Effects related to local  density can be
especially difficult to quantify because automated sample selection can
often miss close neighbours \citep{Verley2007a}. 

 New  imaging and spectroscopic surveys showed that the isolation of
 galaxies is challenged by several factors. The SLOAN Digital Sky Survey
 \citep[SDSS,][and references therein]{Aguado2019} largely contributed both
 to show the presence of faint intruders that can undermine the
 isolation criteria, and, with new redshift measures,   that these
 companions are likely physically associated to the galaxy deemed
 isolated. Statistical considerations (Section \ref{isol}) suggest  that truly isolated galaxies
 could have been isolated for at least 3\,Gyrs, certainly a significant
 fraction of their entire lifetime \citep{Verley2007a}.

Long lasting signatures of past interaction and merging events are still
detectable on the isolated galaxy  structures. A large fraction
($\approx$50-60\%) of  isolated early-type galaxies (iETGs) show a large
variety of such structures like shells,  ripples, streams and  tails
\citep{Hernandez2007, Rampazzo2020a}. Isolated late-type galaxies  (iLTGs) may show
different levels of perturbation in their properties \citep[see
e.g.][]{Fernandez2014, Jones2018}.

In this paper we present a study of five isolated galaxies (Section~2),  for which we obtained
new {\tt Swift-UVOT} observations (PI Trinchieri).  Four of these galaxies are iETGs with good confidence (Section \ref{morpho}). The far-UV (FUV) study of iETGs is {\it per se}
a challenge  since the true (or presumed) isolation together with  the early
morphological type should be  a guarantee of an  extinguished, red and
dead, galaxy. Far-UV observations offer a snapshot  of the recent
(10$^6$-10$^7$ years old) SF activity in galaxies
\citep{Kennicutt2012}. This is relevant also for ETGs. Several ETGs show
 a residual SF activity even if they have reached the red
sequence (RS), as shown by \citet[][]{Jeong2009} and \citet{Marino2011}
studies with {\it GALaxy evolution EXplorer} ({\tt GALEX} hereafter;
\citet{Martin2007}) data.
We recently examined 11 ETGs with SWIFT-UVOT and noticed that the S\'ersic indices in UV are typically lower than optical ones, suggesting the emergence of a disk structure in UV \citep{Rampazzo2017}. This is an indication of a dissipative formation of ETGs, not strictly related to their classification (E vs. S0$/$SA$0$s).
We deepen the study of these galaxies giving insight into  their
evolution  making use of a grid of Smoothed Particle Hydrodynamics
simulations with chemo-photometric implementation (SPH-CPI hereafter) anchored, {\it a posteriori}, to the global properties of our targets.
 We already applied this approach to study ETGs and LTGs evolution in several environments
\citep{Mazzei2014a,Mazzei2014b,Buson2015,Plana2017,Mazzei2018,Mazzei2019}.
The plan of the paper is the following.   In Section 2 we present the
general properties of the five isolated galaxies observed, in Section 3
we detail observations and the reduction techniques adopted.
Difficulties in an accurate surface brightness analysis of UVOT data are
reviewed as well as the reduction packages used. Results, presented in
Section 4, include the UV and the optical surface  brightness
distributions and the comparison of  the optical integrated magnitudes
with those in the literature.  These new data allow us to improve the multi-wavelength spectral energy distributions 
(SEDs) of our targets extending it over about 3 orders of magnitude in wavelength, starting from 0.15 micron.
 Additional information from the literature of our targets are reported in Appendix.
We use  all these data  in Section 5 to constrain, {\it a posteriori}, our simulations. 
In this Section we summarize the recipes of our  SPH-CPI simulations, detailed in previous papers, and present the results of
the simulations which  best match the global properties  of our
targets so giving insight into their evolution. The discussion of
evolutionary properties derived is given in Section 6 and, finally,
in Section 7 there are our conclusions.

Hereafter we use the same cosmological parameters as in \citet{Mazzei2019}:
H$_0$=67\,km\,s$^{-1}$\,Mpc$^{-1}$, $\Omega_{\Lambda}$=0.68, $\Omega_{m}$
=0.32 \citep{Planck Collaboration XVI, Calabrese2017} which
correspond to a $\Lambda$ cold dark matter ($\Lambda$CDM) model with
Universe age of 13.81 Gyr \citep[][see their Table 1]{Calabrese2017},
and a light travel time (look-back time) of 7.98 Gyr at z=1 and 5.23 Gyr
at z=0.5.
 In order to help the  interested reader, we collect in  Table \ref{table0} the acronyms commonly used  through  this paper.

\section{The sample}
The iETGs we are analyzing cover the whole range of morphological classification of ETGs.
In the following we report some considerations about the degree of
isolation and the morphology of our targets whose general 
properties are reported in Table \ref{table1}.
\begin{table*}
\scriptsize
\centering
\caption{General properties of the sample}
\begin{tabular}{llccccccccc}
\hline\hline
Galaxy ID &D$_{3K}$ &Morph. & T  &D$_{25}$   &B  & A$_B$&M$_B$ &M$_{HI}$ & $\eta_K$ & Q$_K$\\
                       &[Mpc]  &   &   &[arcmin]           &[mag]&[mag]   & [mag]    & 10$^9$ [M$_\odot$]  &    &   \\
\hline
CIG~189  &  48.6$\pm$3.4 &   E  &-4.8$\pm$0.8  & 2.09   & 14.19$\pm$0.25 & 0.28 & -19.52$\pm$0.43  &  \dots    & 0.403 & -3.067\\
CIG~309  & 24.8$\pm$1.7 &  Sab(R) &1.6$\pm$0.6 & 4.27 & 11.14$\pm$0.10 & 0.19 &  -21.02$\pm$0.37 &  0.46c    &\dots &  \dots\\
CIG~389  &  25.3$\pm$1.8 & S0B &-1.5$\pm$0.9  &2.34   & 12.86$\pm$0.08 & 0.16 &  -19.32$\pm$0.36 & $<$0.015a &\dots&\dots\\
CIG~481  &  24.2$\pm$1.7 & S0-a&-0.1$\pm$0.5  &2.19      & 13.25$\pm$0.13 & 0.04 &  -18.71$\pm$0.36  & 0.31b &  1.080&  -2.279\\
CIG~637    & 33.5$\pm$2.3 & E-S0&-3.2$\pm$1.0 & 2.40    & 12.59$\pm$0.05 & 0.05 &  -20.09$\pm$0.35 &$<$0.032a&1.132 &-1.432\\
\hline
\end{tabular}
\label{table1}
\tablecomments{  The adopted distances (col.2) are from NED, the morphology (col. 3), the morphological type, T (col. 4), 
projected major axis at the isophotal level
25 mag\,arcsec$^{-1}$ in the B-band, D$_{25}$, (col. 5), 
apparent magnitude in the B-band (col.6),  and our own Galaxy extinction, A$_B$ (col.7), are
from {\tt Hyperleda}. The B-band absolute magnitudes corrected for
Galaxy extinction  are in col.8. The HI masses (col. 9) account for the
distances in col.2 and 1.4 GHz fluxes from NED: $^a$ \citep{Serra2012};$^b$
\citep{Courtois2015}, $^c$ \citep{Haynes2018};  $\eta_k$  (col.10),  the
number density of  neighbor galaxies brighter than M$_B$= -16 mag, and Q$_k$ (col.11), the tidal
strength that these latter produce on our target,  are from
\citep{Verley2007b}.}
\end{table*}
\begin{table}
\centering
\caption{Total exposure times in the {\tt UVOT} filters}
\scriptsize
 \begin{tabular}{llccccc}
\hline\hline
Galaxy & W2  &  M2  & W1   & U & B & V  \\
    ID   & [s]     & [s]    &[s]      &[s]& [s] & [s] \\ 
\hline
CIG~189 &  31165  & 21766 & 16303 &  8016  & 7892  & 7518 \\
CIG~309 &  11196  & 8028   &  5857  &  2638  & 2638  & 2639 \\
CIG~389 &  22219  & 15317 & 11423  & 5701  & 5704  & 5501 \\
CIG~481 &  26741 & 19803  & 13193  & 6572  & 6499  & 6206 \\
CIG~637 &  17620 & 12789  & 9598    &  4601 &  4451 & 4266 \\
\hline
\end{tabular}
\label{table2}
\end{table}
\begin{figure*}
  \centering
  {\includegraphics[width=5.5cm]{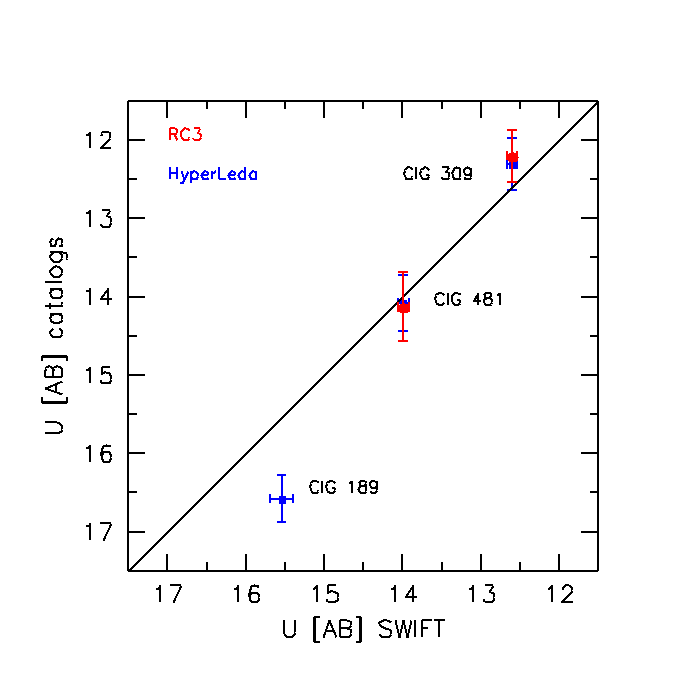}}
 {\includegraphics[width=5.5cm]{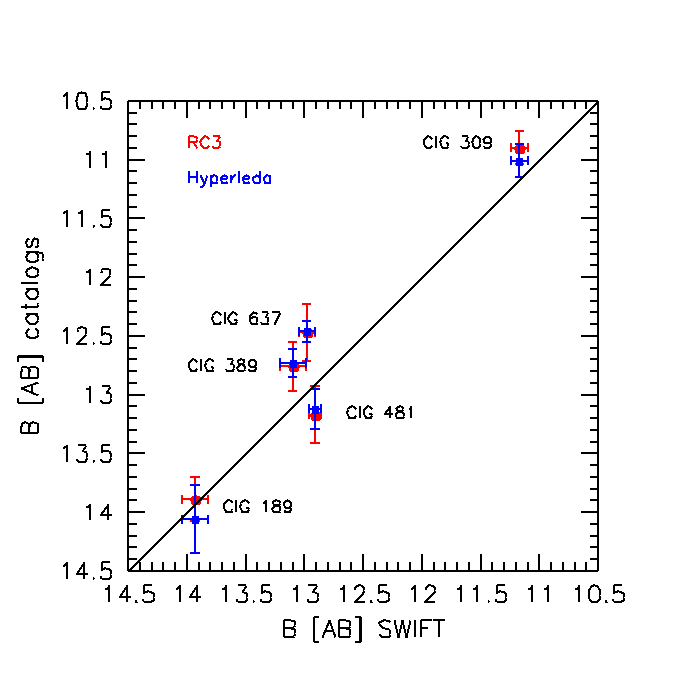}}
  {\includegraphics[width=5.5cm]{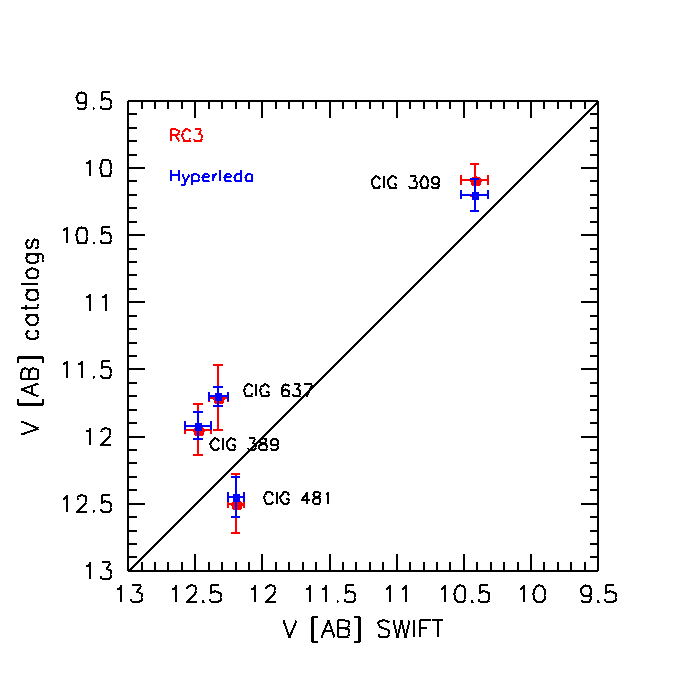}}
      \caption{Comparison between our U, B, and V  [AB] mag and the same in the literature: 
      red from RC3 and blue from {\tt HyperLeda} catalogs}
       \label{fig2}
   \end{figure*}
\begin{figure*}
  \centering
      {\includegraphics[width=14cm]{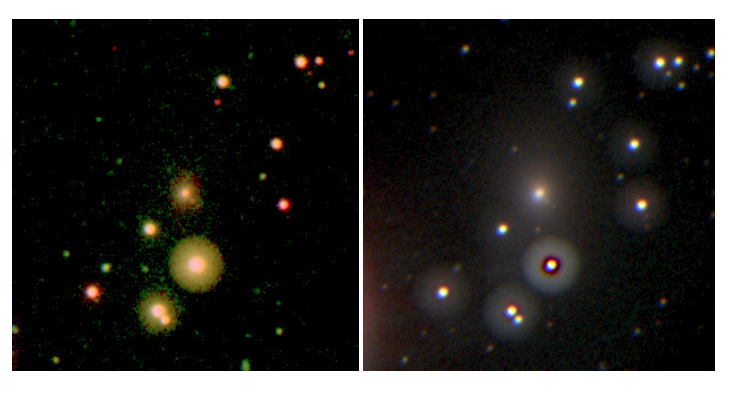}}
     {\includegraphics[width=7.9cm]{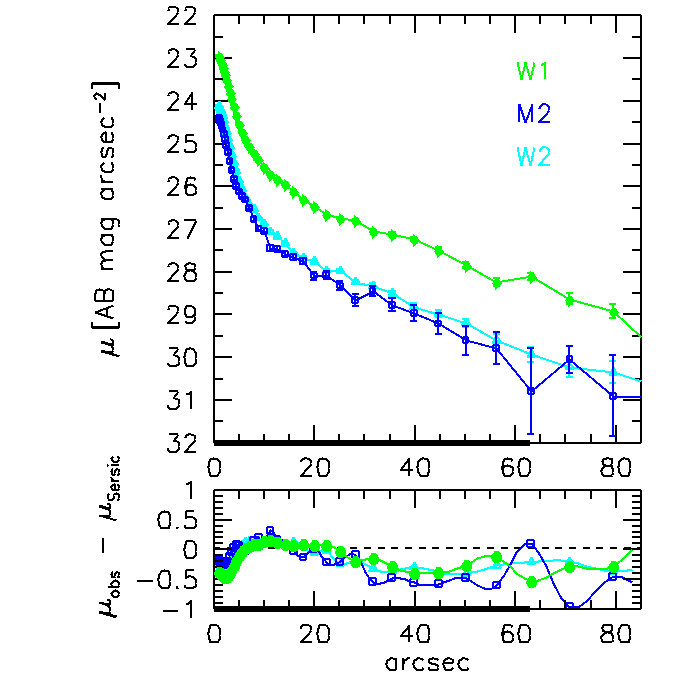}}
      {\includegraphics[width=7.9cm]{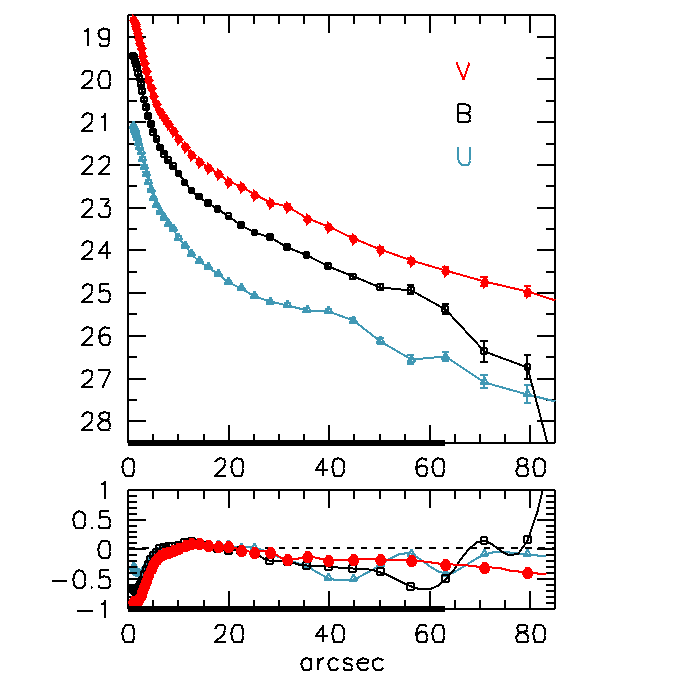}}
       \caption{{\it Top}. Color-composite (left: W2 = blue, M2 = green,  W1=red, right: 
U = blue, B = green, V = red) images of CIG 189 in the {\tt UVOT}
bands. The field of view is
5\arcmin$\times$5\arcmin with North  at the top and East to the left. {\it
Middle}:  Luminosity profiles in the UVOT filters. Bold x-axis line highlights the length of R$_{25}$
in Table \ref{table1}. {\it
Bottom}: The residuals from the best-fit with a single S\'ersic  law (Table \ref{table5}) of the luminosity profiles above accounting for the PSF 
values in Section \ref{obseredu}.} 
\label{fig3}
   \end{figure*}
\begin{figure*}
  \centering
{\includegraphics[width=14cm]{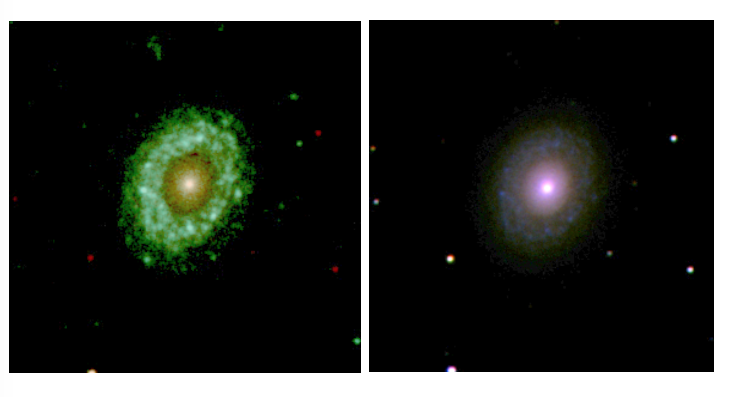}}
      {\includegraphics[width=7.9cm]{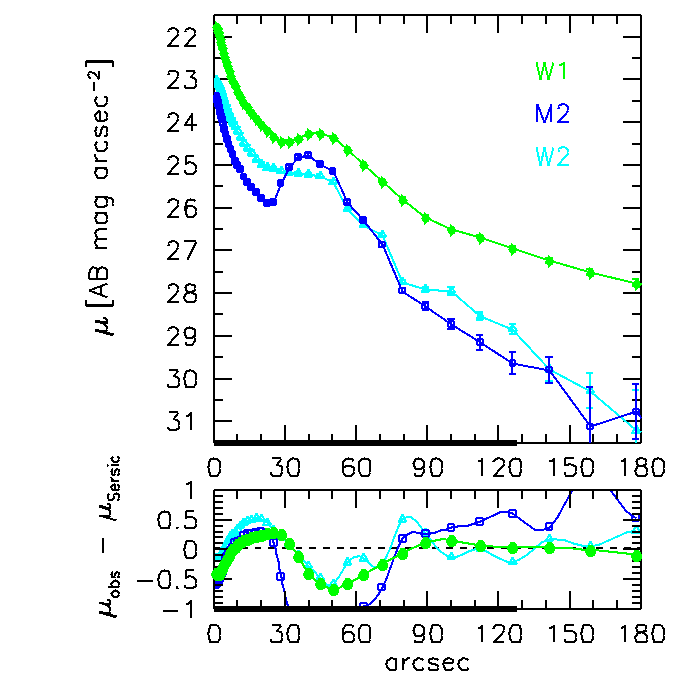}}
        {\includegraphics[width=7.9cm]{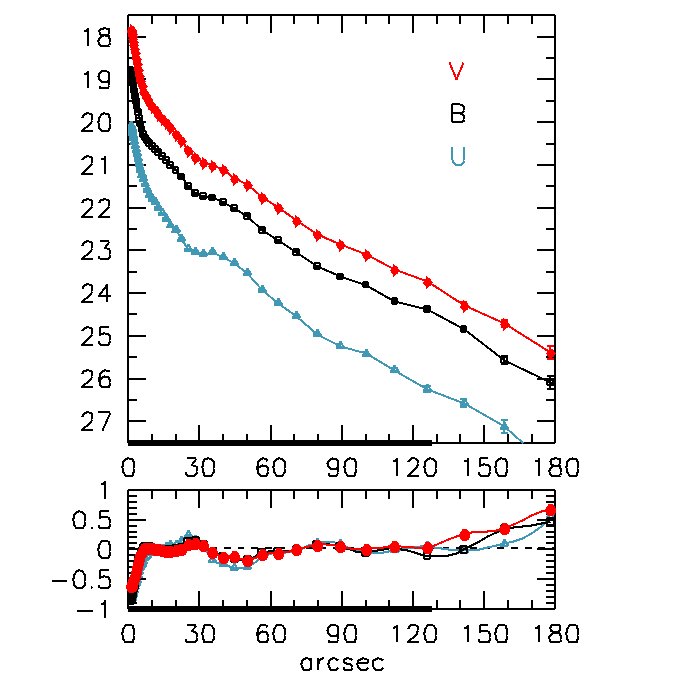}}

      \caption{As in Figure~\ref{fig3} for CIG 309.}
       \label{fig4}
   \end{figure*}
\begin{figure*}
  \centering
 {\includegraphics[width=14cm]{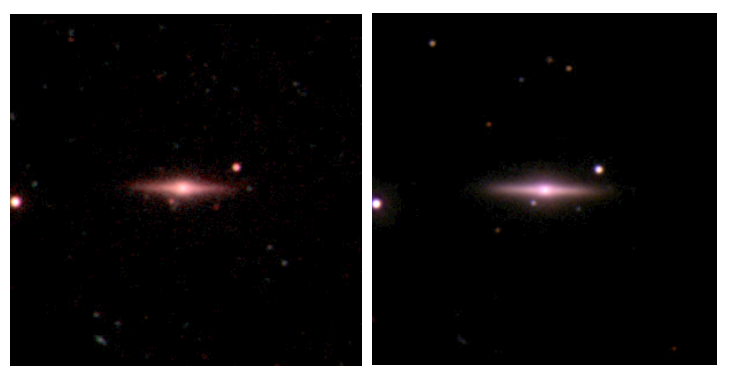}}
      {\includegraphics[width=7.9cm]{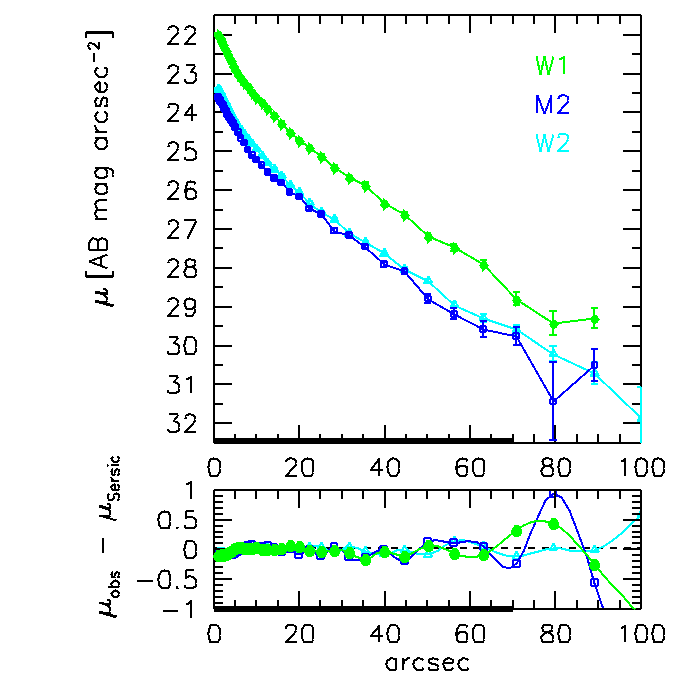}}
        {\includegraphics[width=7.9cm]{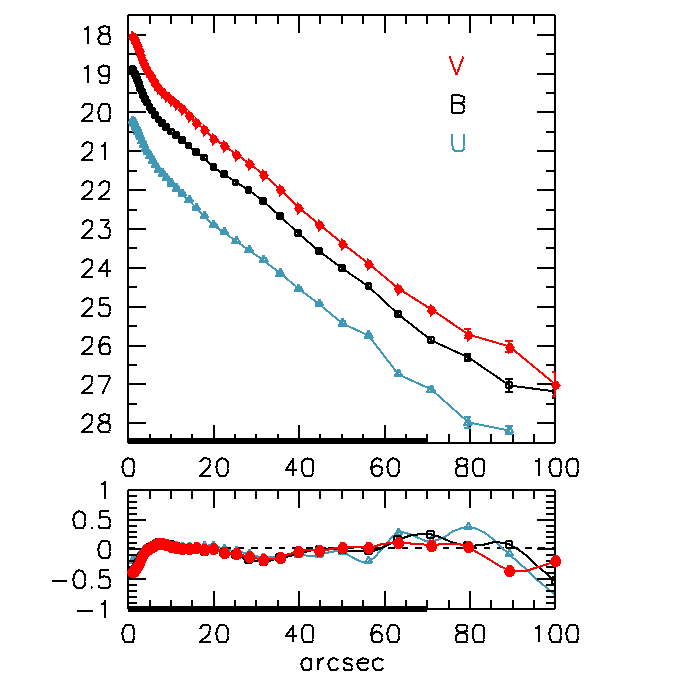}}
      \caption{As in Figure~\ref{fig3} for CIG 389.}
       \label{fig5}
   \end{figure*}
\begin{figure*}
  \centering
       {\includegraphics[width=14cm]{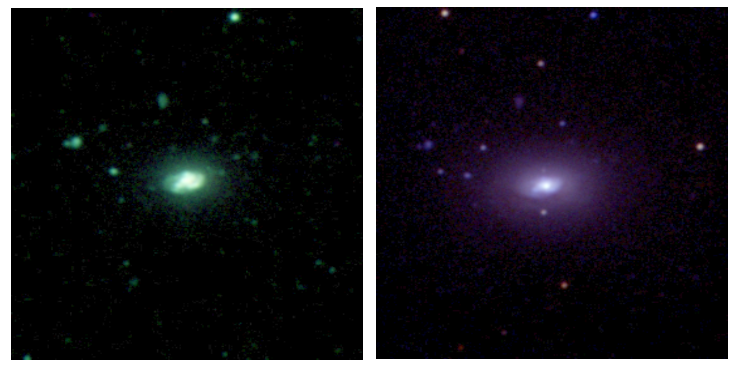}}
      {\includegraphics[width=7.9cm]{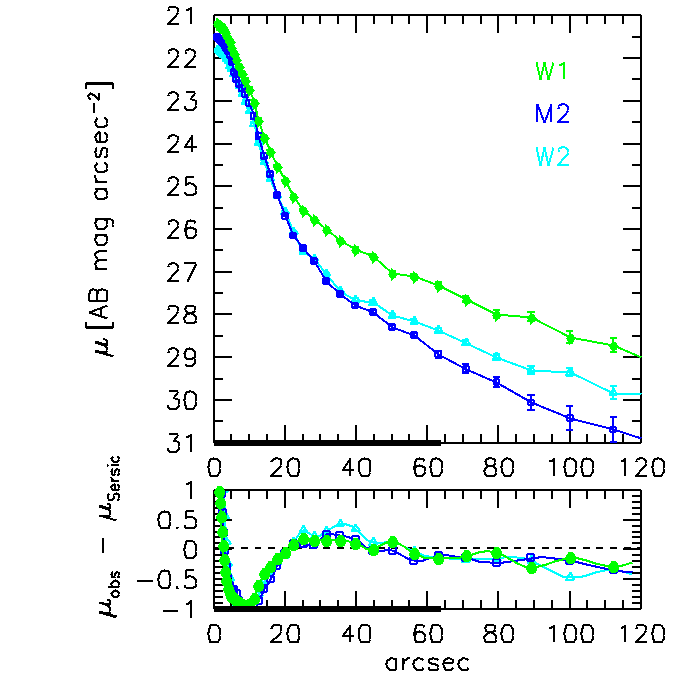}}
        {\includegraphics[width=7.9cm]{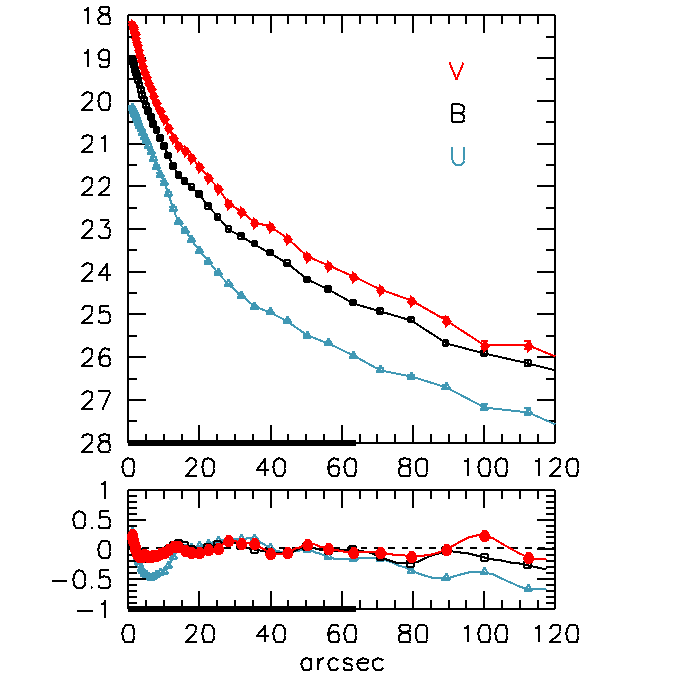}}
      \caption{As in Figure~\ref{fig3} for CIG 481.}
       \label{fig6}
   \end{figure*}

\begin{figure*}
  \centering
         {\includegraphics[width=14cm]{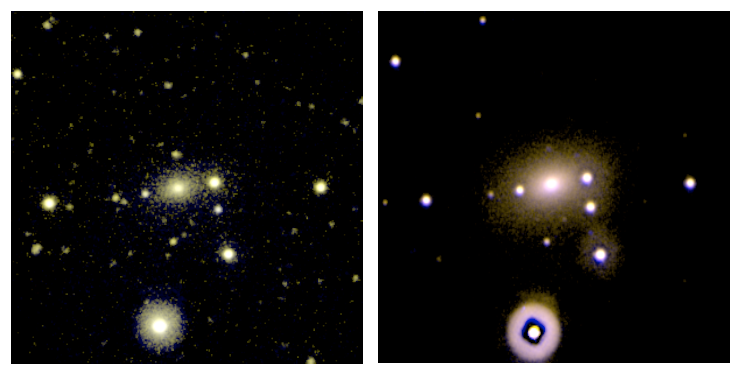}}
      {\includegraphics[width=7.9cm]{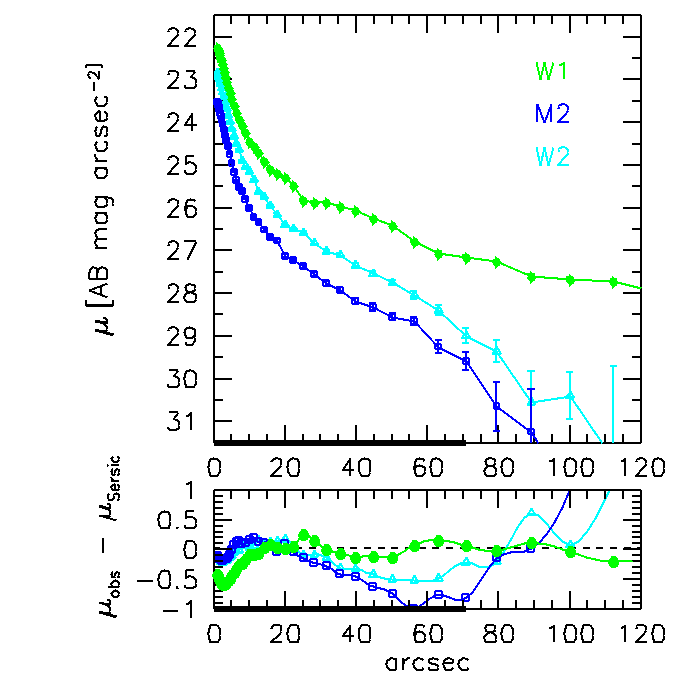}}
        {\includegraphics[width=7.9cm]{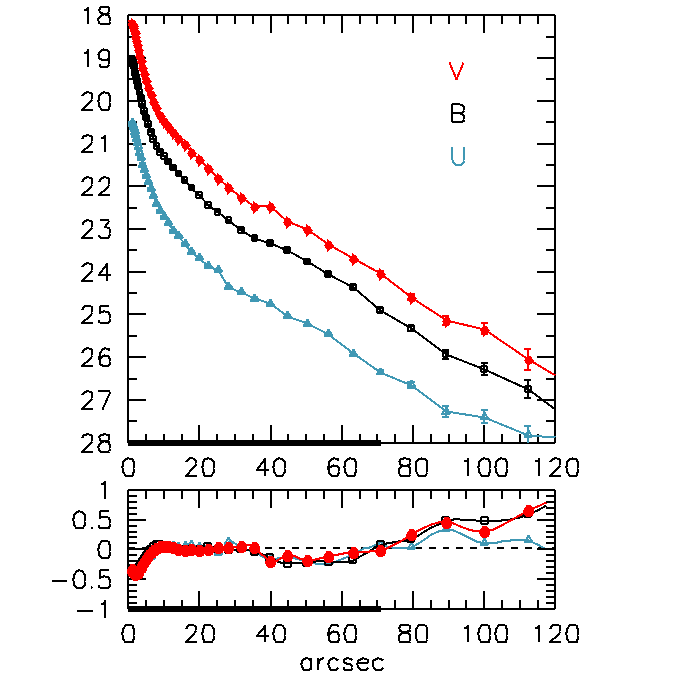}}

      \caption{As in Figure~\ref{fig3} for CIG 637.}
       \label{fig7}
   \end{figure*}
\subsection{Isolation}
\label{isol}
The {\it Catalog of Isolated Galaxies} (CIG) \citep{Karachentseva1973}  
includes 1050 galaxies. The CIG sample was assembled with the 
requirement that no similar sized galaxies, that is neighbors with
angular diameter $d$ between 1/4 and 4 times the angular diameter $D$  of
the primary galaxy,  lie at an angular projected separation from the
primary galaxy smaller than 20$d$.  For  instance, a CIG galaxy with D =
3\arcmin, has no neighbor with d= 12\arcmin\   within 240\arcmin\ and
no companion with d = 0.75\arcmin   within 15\arcmin.  However, 
dwarf companions are not excluded.   Therefore, if one assumes an
average D = 25 kpc for a CIG galaxy and a typical "field" velocity of
150\,km~s$^{-1}$ then an approximately equal mass perturber (d=D) would
require 3.2$\times$10$^{9}$ years to traverse a distance of 20d
\citep{Stoke1978, Verley2007a}.  This suggests that the galaxy did not suffer strong
interactions in the last 3.2\,Gyrs. 

Here we are analyzing five CIG
galaxies with new {\it SWIFT} {\tt UVOT} observations. Two of them, CIG~309, and CIG~389,  are
nearby galaxies with a recession velocity V$_{hel}<$1500 km~s$^{-1}$, 
so are not included in the reassessment of the isolation degree of CIG
galaxies by \citet{Verley2007a}, neither in the following 
quantification of the degree of isolation  \citep{Verley2007b}. 
CIG 309 and CIG 389 are in the outermost parts of the local
supercluster \citep{Verdes2005}.  As noted by \citet{Haynes1983} in
spite of the fact that most of the isolated galaxies are outer
components of groups or clusters, so that it seems difficult to find a
truly isolated population of galaxies, however  it is possible to have
access to regions of very low galaxy density, where the galaxies reflect
properties characterizing their formation. CIG~309,  also known as NGC~2775, as
an example,  is a member of the same group as NGC~2777
\citep{Shapley2001}, LGG~169 \citep{Garcia1993} which includes 3 member
galaxies. Galaxies in the CIG have been found to show different degrees
of isolation \citep{Verley2007a,Verley2007b}.   CIG~189, CIG~418, and
CIG~637,  actually lie in very low density environments given that their
local number density parameter, $\eta_k$ (Table 1), is below the
critical value, 2.4. Moreover, the tidal strength estimation, Q$_k$,  is
lower than the critical value, -2,  for CIG~189 and CIG~481 (Table 1)
which are isolated from any sort of external influence. The same is not true for
CIG~637 which lies in a very low density environment but can suffer the
consequences of tidal interactions with some dwarf companion.

\subsection{Morphologies}
\label{morpho}

In addition to a reassessment of their isolation, several galaxies in
CIG have been also re-classified many times \citep[see
e.g.][]{Fernandez2012,Buta2019,Rampazzo2020a}.
CIG~309 deserves a special mention. {\tt HyperLeda}
classified this galaxy as a Spiral with an outer  ring
(Table~\ref{table1}). BVRI photometry reveals a ring enclosing a "soft"
bulge  \citep{Hernandez08}. The spatial distribution of the inner star
forming regions and dust lanes traces the arms. For \citet{Comeron2014} 
NGC~2775 is a SA(rl)0$^+$, that is a lenticular galaxy with a inner
resonance ring. Indeed this galaxy does not present arms  and the disk
extends well outside the ring   (see their Appendix B1). The  
classification suggested by  near-IR data of \citet{Comeron2014}  is well supported by the high
stellar circular velocity, 270$\pm$4\, km\,s${^{-1}}$, and
central velocity dispersion, 162$\pm$13\,km\,s${^{-1}}$, of
\citet{Ferrarese2002}. \citet[][upper panel in their Fig.8]{Morales18}  do
not find any signature of spiral arms. However, they find a tidal stream
extending 29\,kpc around this galaxy which they define as a shell.
Shells are faint structures found in ETGs as remnants of a merging
episode \citep{MC83,Dupraz1987} especially in ETGs in low density
environments where mergers are likely to occur
\citep{Reduzzi1996,Rampazzo2020a}. We conclude that the morphological
properties of CIG~309 are likely of an ETG at the border with a LTG.
.

CIG~389 (NGC~3098) is a barred lenticular galaxy according to {\tt
HyperLeda} (Table~\ref{table1}). This  classification has been revised
by \citet{Buta2015} in SAB:0$^- sp/E8$ that corresponds to a value 0.25
of the type (T) parameter of {\tt HyperLeda}. \citet{Laurikainen2017} found
that the galaxy has a very compact X-structure in the centre \citep[see
also][]{Yildiz2020}.

Concentric shells are  shown by CIG~481  \citep{Morales18},
a peculiar S0  discussed in \citet{Rampazzo2020a}. 
They also support   a S0 morphology of CIG~637
which however reveals a inner and outer ring in the residuals of a  S\'ersic fit profile subtraction.

\section{Observations and data reduction}
\label{obseredu}

UVOT is a 30 cm telescope in the {\it Swift} platform operating both in
imaging and spectroscopy \citep{Roming2005}. We observed our targets in
imaging in all six available filters, W2 ($\lambda_0$ 2030 \AA), M2
($\lambda_0$ 2231 \AA), W1 ($\lambda_0$ 2634 \AA),  in the ultraviolet
(UV) spectral range, and U ($\lambda_0$ 3501 \AA), B ($\lambda_0$ 4329
\AA), V ($\lambda_0$ 5402 \AA) in the optical one. Description of the 
filters,  their PSFs (2.92\arcsec\ for W2, 2.45\arcsec\ for M2,
2.37\arcsec\ for W1, 2.37\arcsec\ for U, 2.19\arcsec\ for B,
2.18\arcsec\ arcsec for V), and calibrations are discussed in
\citep{Breveeld2010,Breveeld2011}. UVOT data obtained in imaging mode
with a 2$\times$2 binning, resulting in 1.004\arcsec/pixel of
resolution, were processed using the procedure described
online\footnote{http://www.swift.ac.uk/analysis/uvot/}. We combined all
the images taken in the same filter for each galaxy in a single image
using {\tt UVOTSUM}, binned 2$\times$2 to improve the S/N and to enhance
the visibility of UV features of low surface brightness. The final
exposure times per image are different since we complied with the
request of preserving the lifetime of the filter wheel and therefore
observed as much as possible in the filter-of-the-day. Total exposure
times are reported in Table 2.  We used the photometric zero points
provided by \citet{Breveeld2011} for converting UVOT count rates to the
AB magnitude system V = 17.88$\pm$0.01, B = 18.98$\pm$0.02, U =
19.36$\pm$0.02, W1 = 18.95$\pm$0.03, M2 =18.54$\pm$0.03 and W2 =
19.11$\pm$0.03 \citep{Oke1974}.

UVOT is a photon counting instrument and is therefore subject to
coincidence loss when the throughput is high, whether this is due to
background or source counts, which may result in an undercounting of the
flux. This effect is a function of brightness of the source and affects
the linearity of the detector. The U, B, and V filters are most
affected, although coincidence loss can also be present in UV filters.
For our binning, \citet[][and reference therein]{Hoversten2011}
estimate that count rates lower than 0.028 counts s$^{-1}$
pixel$^{-1}$ are affected by at most 1\% due to coincidence loss. We
checked for coincidence loss in all our images. The UV filters W2,
M2, and W1 are almost free of coincidence loss effects. We verified
that in W1, which is the most affected of the UV filters, the region with count rate $>$ 0.028 counts s$^{-1}$
pixel$^{-1}$
is restricted to a few pixels centered on the galaxy nucleus. For
instance, in the case of CIG 637, the level of the 0.028 counts s$^{-1}$
pixel$^{-1}$ in the W1 filter is exceeded in four central pixels.
Coincidence loss effects can be corrected in the case of point sources
\citep{Poole2008,Breveeld2010}. For extended sources a correction
process has been performed for NGC 4449, a Magellanic-type irregular
galaxy with bright star-forming regions, by \citet{Karczewski2013}. Even
though their whole field  is affected, the authors calculated that the statistical and systematic uncertainties in their total fluxes amount to
7-9\% overall for the UV and the optical bands. Based on these results,
we decided to refrain from applying any correction for coincidence loss
to our optical and UV data.

We considered the presence of instrumental scattered light that, in the
UV filters, may cover the whole frame and of the light scattered from
stars \citep[see e.g.][]{Hodges2014}. This latter may produce
ghost-images: for particularly bright stars, a typical ring pattern is
produced. The most contaminated frames are those of CIG 637, in the W1
filter (see next section). However, our images are the sum of several
dithered and rotated frames. This sum tends to smooth out large-scale
inhomogeneities in the final frame, while ghosts of bright stars remain
a serious problem for an accurate surface photometry. Our targets,
however, cover a limited portion of the 17\arcmin$\times$17\arcmin of
the {\tt UVOT} field of view. The background, although not homogeneous
due to the above factors, can be well evaluated around each object. The
surface photometry has been performed using the {\tt ELLIPSE}
\citep{Jedrzejewski1987} fitting routine in the STSDAS package of IRAF.
Foreground and background objects have been removed substituting the
surrounding background using the IRAF task {\tt IMEDIT}. Ghosts, very
bright and extended only in the case cited above, have been masked using
{\tt ELLIPSE}. To secure a reliable background measure, we forced the
measure well beyond the galaxy emission. 
From the surface brightness profiles, we derived apparent magnitudes by
integrating the surface brightness within elliptical isophotes. Errors
of the magnitudes where estimated by propagating the statistical errors
on the isophotal intensity provided by {\tt ELLIPSE}. Our integrated
magnitudes, reported in Table~\ref{table3}, are not corrected for
Galactic extinction.

\begin{table*}
\centering
\scriptsize
\caption{UV and optical integrated magnitudes in the AB system}
\begin{tabular}{lcccccc}
\hline\hline
Galaxy  & W2      & M2   & W1     & U       &    B     & V \\
ID         & [mag]  &[mag] & [mag] & [mag] & [mag]  & [mag] \\
\hline\hline
CIG~189 &18.74$\pm$0.07 & 18.74$\pm$0.14 & 17.31$\pm$0.10 & 15.54$\pm$0.15 & 13.93$\pm$0.11 &12.98$\pm$0.10 \\
CIG~309 &15.04$\pm$0.05 &15.10$\pm$0.12 &13.93$\pm$0.06 &12.60$\pm$0.06 & 11.17$\pm$0.07 &10.42$\pm$0.10 \\
CIG~389 &17.66$\pm$0.14 & 17.67$\pm$0.19 &16.26$\pm$0.07 & 14.47$\pm$0.15 &13.10$\pm$0.11 &12.48$\pm$0.10 \\
CIG~481 & 15.84$\pm$0.04 &15.78$\pm$0.08 & 15.09$\pm$0.08 &13.99$\pm$0.07  & 12.91$\pm$0.05 &12.20$\pm$0.06 \\
CIG~637 &17.61$\pm$0.07 & 17.78$\pm$0.14 & 15.86$\pm$0.13 & 14.48$\pm$0.11 & 12.98$\pm$0.07 &12.33$\pm$0.07 \\
\hline
\end{tabular}
\label{table3}

\tablecomments{Magnitudes  are not corrected for  extinction from our own Galaxy}

\end{table*}

\begin{table*}
\centering
\scriptsize
\caption{Comparison of B integrated magnitudes with the literature}
\begin{tabular}{lcccc}
\hline\hline
Galaxy  & B$_T$           & B$_T$   & B$_T$                       & B$_T$\\
ID         & [HyperLeda]  &[NED]     & [Verdes-Montenegro] & [our] \\
\hline
CIG~189  &   14.19$\pm$0.25  &   14.02$\pm$0.15  &   14.30$\pm$0.14  & 14.06$\pm$0.11\\
CIG~309  &   11.14$\pm$0.10  &   11.03$\pm$0.10  &   11.40$\pm$0.14  & 11.30$\pm$0.11 \\
CIG~389  &   12.86$\pm$0.08  &   12.89$\pm$0.17   &  13.00$\pm$0.14  & 13.23$\pm$0.11 \\
CIG~481  &   13.25$\pm$0.13  &   13.42$\pm$0.18   &  13.40$\pm$0.14  & 13.04$\pm$0.05 \\
CIG~637  &   12.59$\pm$0.05  &   12.60$\pm$0.20   &  13.30$\pm$0.14  & 12.98$\pm$0.07  \\  
\hline
\end{tabular}
\label{table4}

\tablecomments{ All   magnitudes are not corrected for extinction from our own Galaxy; 
col.5  accounts for  the m$_{AB}$-m$_{Vega}$ conversion, -0.13 (Section \ref{resmag})}

\end{table*}

\begin{table*}
\centering
\scriptsize
\caption{Indices of the single S\'ersic fit}
\begin{tabular}{lcccccc}
\hline\hline
 Galaxy  & W2      & M2   & W1     & U       &    B     & V \\
ID         &           &         &            &           &       &  \\
\hline
CIG~189 &  3.23$\pm$0.18 & 2.79$\pm$0.10 & 2.85$\pm$0.12 & 3.19$\pm$0.19 &2.36$\pm$0.07 &2.43$\pm$0.11 \\
CIG~309 &  1.29$\pm$0.03 & 1.70$\pm$0.13 & 2.07$\pm$0.07 &1.65$\pm$0.01 &1.74$\pm$0.03 &2.10$\pm$0.07 \\
CIG~389  & 1.65$\pm$0.04 & 1.58$\pm$0.04 & 1.56$\pm$0.04 &1.37$\pm$0.01 &1.24$\pm$0.03 &1.26$\pm$0.03 \\
CIG~481  & 5.12$\pm$0.13 & 4.23$\pm$0.22  &4.61$\pm$0.24 &2.77$\pm$0.01 &2.87$\pm$0.03 & 2.97$\pm$0.05 \\
CIG~637  & 2.52$\pm$0.07 & 2.30$\pm$0.15 &3.67$\pm$0.14 & 2.37$\pm$0.07 & 2.54$\pm$0.08 &2.49$\pm$0.01 \\
\hline
\end{tabular}
\label{table5}
\end{table*}

\section{ Photometry results}
\label{resobs}

\subsection{{\tt UVOT} Integrated magnitudes}
\label{resmag}
Table~\ref{table3}  shows the integrated magnitudes we derived.
We compare our AB integrated optical (U, B and V)  magnitudes with RC3 
\citep{deVaucouleurs1991} and {\tt Hyperleda} catalogs in Figure \ref{fig2}   by applying the  
following corrections,  AB-Vega: V = -0.01, B = -0.13, U = 0.79, as in  \citet{Rampazzo2017}.
Table~\ref{table4} extends the comparison of our B integrated magnitudes  
with other available data in the literature.
Most of our  magnitudes are consistent, within errors, with those in the 
previous catalogs indicating that the effects discussed above (Section 3) do not 
affect our measures significantly.  Moreover, B and V magnitudes of CIG~309 agree  well  with the results of 
\citet[][their Table 3]{Hernandez2007}.

New U magnitudes for CIG~389 and CIG~637, and   V magnitude 
for CIG~189 are derived together with  new  UV magnitudes in  three   bands for each of our targets.

\subsection {{\tt UVOT} surface photometry}
\label{surfphot}

Top panels of Figure~\ref{fig3}  to Figure~\ref{fig7} show 
color-composite images of our targets   in the UV (left)  and in the optical bands (right).
Luminosity profiles are presented in AB magnitudes in the middle
panels of the same figures, bottom panels show the fit residuals.

To parameterize the shape of   luminosity profiles, we adopted a
simple S\'ersic law  \citep[$\mu \propto$ r$^{1/n}$;][]{Sersic1968}, which is widely used
for elliptical and S0 galaxies since it is a generalization of the 
r$^{1/4}$ \citet{deVaucouleurs1948}  law  \citep[see
also][]{Caon1993}. Special cases are n = 1, the value for an exponential
profile, and n = 0.5, a Gaussian luminosity profile. Galaxies with n
values higher than 1 have a steep luminosity profile in their nuclear
regions and extended outskirts. Values lower than 1 indicate a flat
nuclear region and more sharply truncated outskirts. From a 2D
luminosity profile decomposition of about 200 Ellipticals from the SDSS,
\citet{Gadotti2009} measured that the S\'ersic index in the i-band has a
mean value of 3.8$\pm$0.9, close to n = 4, the historical paradigm for
bona fide Ellipticals \citep{deVaucouleurs1948}, although with a large
scatter. The multi-wavelength analysis by \citet{LaBarbera2010}  showed
that  n$_i$ values of ETGs  are quite stable, with average values from
5.44$\pm$2.36 in g-band to 6.68$\pm$2.60 in K-band. In the following  we
summarize our photometric results for each target together with 
kinematical properties, if any.  We will use all these data in Section
\ref{Sims} to constrain {\it (a posteriori)} our simulations. We
consider $n=$2.5 as the transition value  between galaxies without
a disk ($n>$2.5) and those with a disk  ($n<$2.5).  
 We perform a 1D best fit with a
S\'ersic law convolved with a point-spread function (PSF) using a custom
IDL routine based on the MPFit  package \citep{Markwardt2009} and
accounting for errors in the surface photometry, as in
\citet{Rampazzo2017}. The PSF model is a Gaussian of a given full width
at half maximum (FWHM), and the convolution is computed using fast
Fourier transformation (FFT) on oversampled vectors. We used the nominal
value of the FWHM of the PSF for each UVOT filter. However,  as a result
of the coadding, binning, and relative rotation of the frames, the FWHM
is broadened by about 15\% compared to the nominal value
\citep{Breveeld2010}. We verified that the effects of the small
variations of the PSF are well within the error associated with the
S\'ersic index.
The derived values of S\'ersic indices, n$_i$,  in the six UVOT filters
are presented in Table~\ref{table5}.
The residuals from the best-fit with the single S\'ersic  laws  in Table \ref{table5} of the luminosity profiles of our targets
are reported in the bottom panels of Figure \ref{fig3} to Figure \ref{fig7}.

{\it CIG 189.} This Elliptical galaxy, UGC03844,is in a
crowded field, both in the optical and in the UV
filters (Figure \ref{fig3}, top). Its luminosity profiles (Figure
\ref{fig3}, middle)  extend beyond R$_{25}$ (Table~\ref{table1} and tick x-axis line) showing small errors,
with the exception of M2 and W1 profiles  at r$\ge$90\arcsec.The average value of the S\'ersic indices in
Table~\ref{table5}  is   2.81$\pm$0.36. We do not find  any significant
difference between the  optical and UV average values,  2.66$\pm$0.46,
and 2.96$\pm$0.24, respectively. 

 {\it CIG 309.} 
This galaxy, alias NGC 2775 or UGC 4820, has an uncertain classification
as explained in Section~\ref{morpho}.
Color-composite  images  and its luminosity profiles are provided in
Figure~\ref{fig4} (top and middle, respectively). Following
\citet{Shapley2001}, this is the brighter member of  a low density group
together with NGC 2777, two magnitude fainter. Its luminosity profiles
extend well beyond R$_{25}$, up to 128\arcsec (Table~\ref{table1}). The
outer ring   is very bright in the UV  and it is  still  well visible in the
optical bands.
This galaxy is an interesting example of  ring formation  in isolated
environment. Most rings form by gas accumulation at resonances, usually
under the continuous action of gravity torques from a bar pattern, but
sometimes in response to a mild tidal interaction with a nearby
companion \citep{Buta1996, Buta1999}. A resonance is a very special
place in a galaxy where SF can be enhanced and may proceed
either as a starburst or continuously over a period of time. Most of the
observed rings in isolated galaxies \citep{Hernandez2007}  show blue
colour distributions, suggesting that their stellar populations are
similar to those in their hosting disks. The composed JHK image  of this
galaxy by \citet[][their Fig. B.9, bottom panel]{Hernandez2007} also
shows the  ring. The existence of an old stellar population underlying
the star-forming one suggests a strong coupling between the stellar and
gaseous components in this resonance region.\\
The average value of the S\'ersic indices in Table~\ref{table5}  is 
1.76$\pm$0.30. There are no  meaning differences between the averages of
optical and UV  values, 1.83$\pm$0.24 and 1.69$\pm$0.39
respectively.  These results indicate the presence of  a stellar disc in the UVOT filters.
The H-band surface brightness profile decomposition  of 
\citet{Fabricius2012} points towards a higher value with, however,  a large 
uncertainty, 3.23$\pm$0.93, in the inner regions of this galaxy. 

{\it CIG 389.} This edge-on  S0 barred galaxy is also known as
NGC 3098.  Its luminosities profiles are in Figure~\ref{fig5}  below 
color-composite UV and optical images. The major axis of this galaxy in the
B-band is about 1\arcmin (RC3). The surface brightness profiles  extend up to 1.5 times this limit.  
The mean value of the 
S\'ersic indices in Table~\ref{table5}  is  1.44$\pm$0.18 with 
1.29$\pm$0.07 for the optical 
and  1.59$\pm$0.05 for UV indices. All these values hint at the presence of
a well defined stellar disc.

{\it CIG 481.} This  galaxy, also known as NGC 3682
(Table~\ref{table1}), has a very uncertain classification being,
probably, a peculiar S0 since does not show arms \citep{Rampazzo2020a}.
\cite{Morales18}   found two classical shells  on both sides of the galaxy.
Figure~\ref{fig6} shows UV and optical color-composite images and the
luminosity profiles. Our profiles extend almost up to
2$\times$R$_{25}$. The average value of the  S\'ersic indices in
Table~\ref{table5}  is 3.76$\pm$1.02, as expected for ETGs. The average
value of UV  indices, 4.65$\pm$0.45, is higher than that of optical ones,
2.87$\pm$0.10 suggesting the presence, in the inner 20\arcsec, of a very asymmetric component, with a complex geometry, not well matched by   1D  UV fits.

 {\it CIG~637.} This nearby galaxy (Table~\ref{table1},  see also
\citet[][]{Rampazzo2020a} and references therein), also known as NGC
5687, belongs to a quite crowded field, with several bright optical
stars projected around the galaxy body. Figure~\ref{fig7} shows its UV and
optical composite images and the luminosity profiles we derived. The
optical brightness profiles extend almost up to 2$\times$R$_{25}$. The average
value of the  S\'ersic indices in Table~\ref{table5}  is 2.64$\pm$0.50,
with the average of UV indices and optical ones, 2.83$\pm$0.74 and
2.47$\pm$0.09 respectively. These values are border-line with the transition value for ETGs (n$=$2.5)
pointing towards a disk component as discussed in Section \ref{morpho}  and references therein.

Summarizing, the results of our photometric study suggest that: 
1)  iETGs in our sample cover the entire range of ETGs, from {\it bona fide} Ellipticals
(CIG 189) to late S0s (CIG 309) borderline with LTGs; 2) this variety is confirmed 
also by S\'ersic indices running from about 1.3 to 4.8 with sometimes significant 
differences between optical and UV \citep[see also][]{Rampazzo2017}; 
3) the mass of  HI gas shows a significant range, from  7.19 to 8.58 in log(M$_{HI}$[M$_{\odot}$]). 
According to \citet{Serra2012}  the HI gas is typically distributed in a small 
disk within the galaxy.

\section{SPH+CPI simulations} 
\label{Sims}
\begin{table*}
\centering
\caption{Initial Parameters of SPH-CPI Simulations reproducing  the 5 CIGs}
\scriptsize
 \begin{tabular}{lcccccccc}
\hline\hline
Galaxy & M$_{tot}$                       & Mass ratio  & N$_{part}$ & r$_1$ & r$_2$ &  v$_1$ & v$_2$ & spins \\
    ID   & [10$^{10}$ M$_\odot$]  &    &                                    & [kpc]   & [kpc]   & [km~s$^{-1}$] & [km~s$^{-1}$] & \\ 
\hline
 CIG~189   &  250      &         5:1      &  20.0$\times$10$^4$    &   93    &  373   &  33   &   131     &      T   \\
 CIG~309*  &  500      &      1:0.5$+$1      &   20.0$\times$10$^4$   &  701,198   & 621    & 121,27    &  108      &    $\updownarrow$  + T \\
CIG~389    &  220      &       10:1       & 17.6 $\times$10$^4$   &   15    & 149   &  14    &  140     &  $ \parallel$            \\
CIG~481     & 200      &       1:1          &   20.0$\times$10$^4$    &  446    & 446   &  71   &   71      &  $\parallel $         \\
CIG~637*   & 250      &        1:1$+$0.5     & 20.0$\times$10$^4$      &  427,359    &   217   & 118, 56    & 66     &  $\parallel$ +  $\updownarrow$   \\
\hline
\end{tabular}

\tablecomments{ Our targets are in col.1, the total mass of the simulation selected in col.2, the halo mass ratio in col.3, 
the initial number of particles in col.4; positions and velocities of the two halos with respect 
 to the mass center of the system in col.s 5-8, the halo spin direction ($\parallel$ parallel; T perpendicular; $\updownarrow$ 
 counter-rotating) in col.9;  * indicates simulations with 3 initial halos. }
\label{table6}
\end{table*}
\begin{figure*}
  \centering
 {\includegraphics[width=15cm]{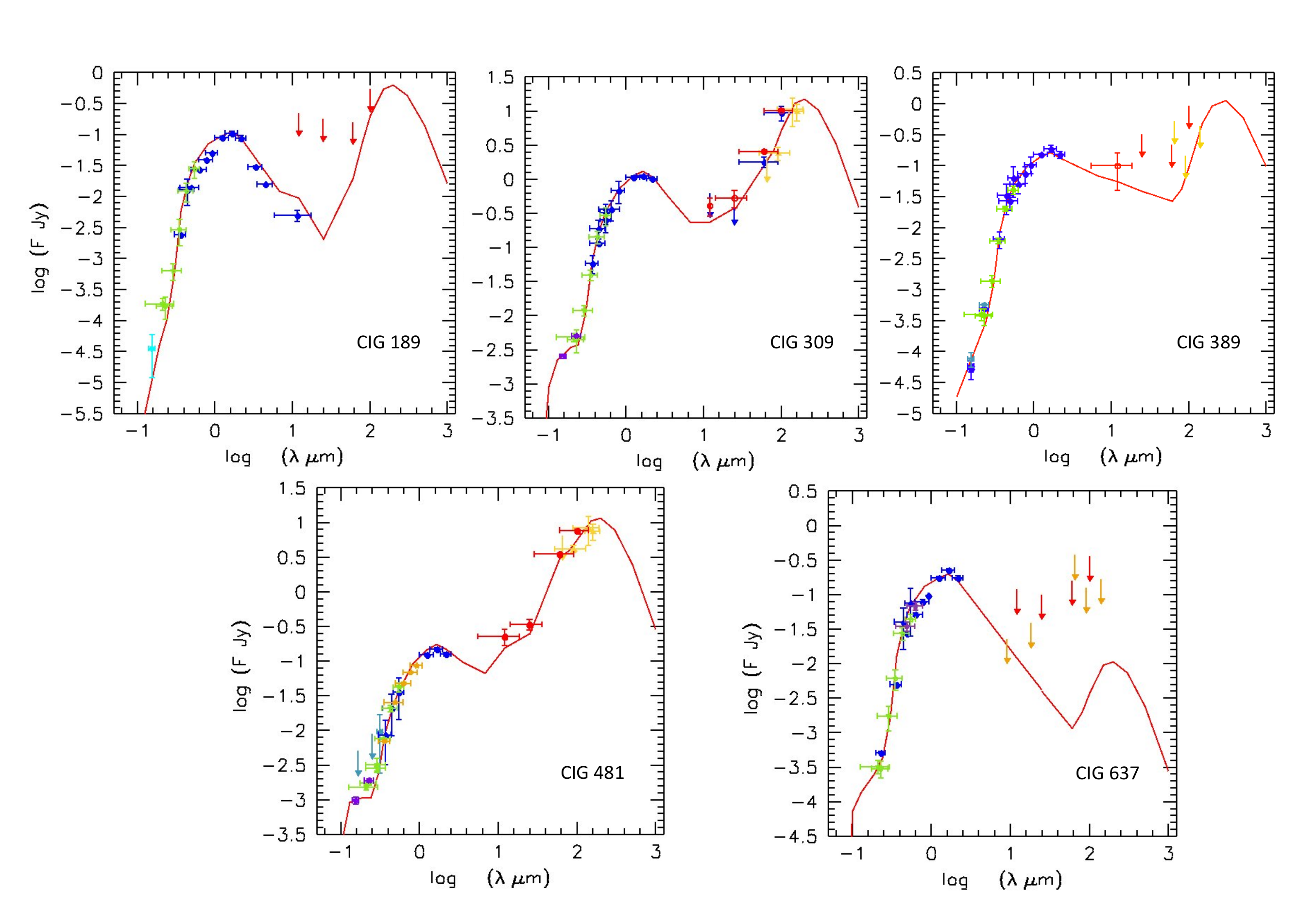}}
\caption{SEDs of our targets (colored dots) compared with predictions from 
SPH-CPI simulations in Table 6 (red solid lines) at the galaxy ages in Table~\ref{table7}. 
Green triangles are Swift-UVOT magnitudes in  this paper, blue filled circles are from 
NED, red points are IRAS fluxes from \citet{Lisenfeld2007}; other details are in the 
subsection focusing on each target  (Section \ref{sim189}--\ref{sim637}).
All the data are corrected for our own Galaxy extinction following prescription of NED 
and Paper II in the Swift-UVOT filters. 
}
       \label{fig8}
   \end{figure*}
\begin{figure*}
  \centering
 {\includegraphics[width=8.97cm]{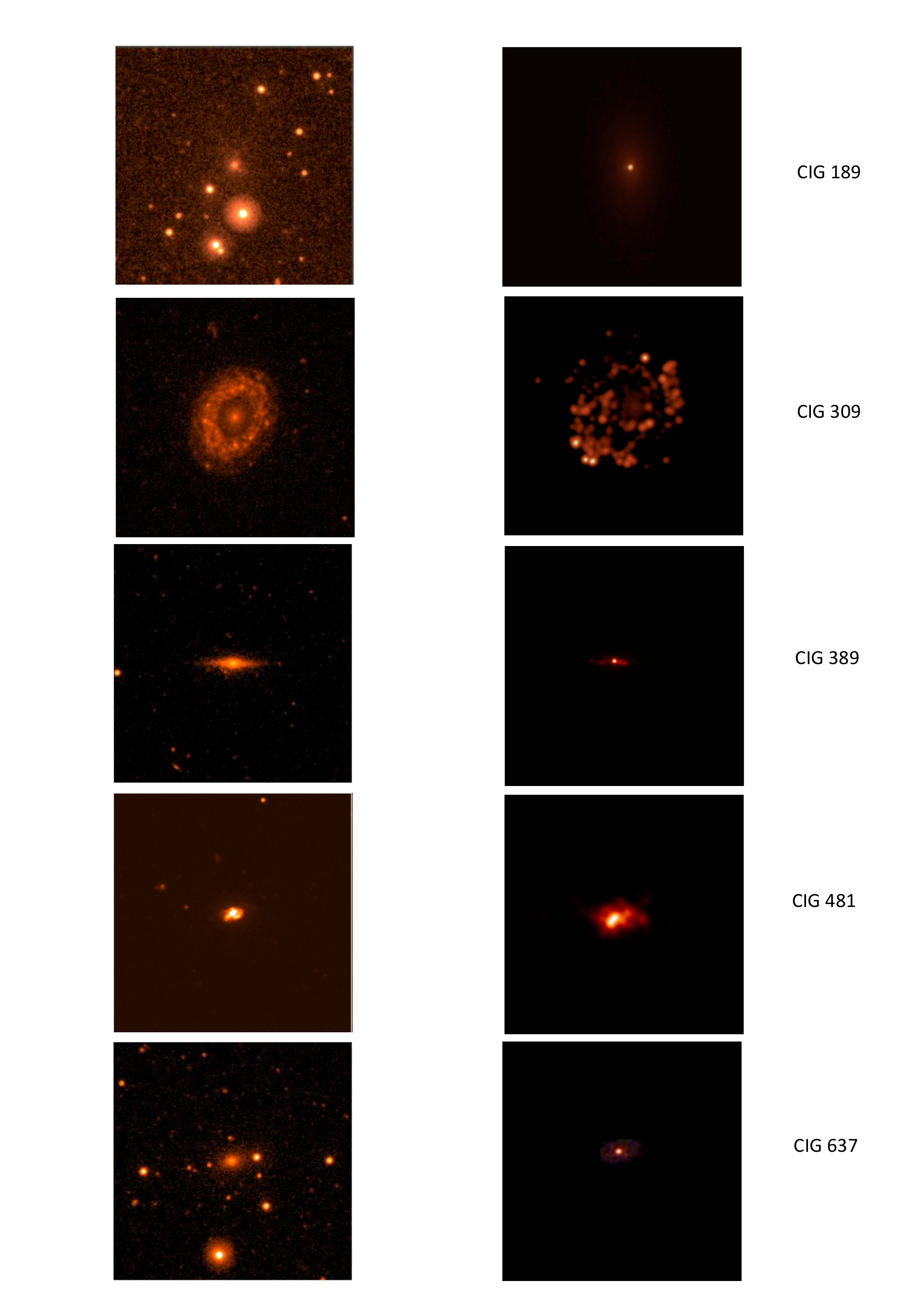}}
 {\includegraphics[width=8.97cm]{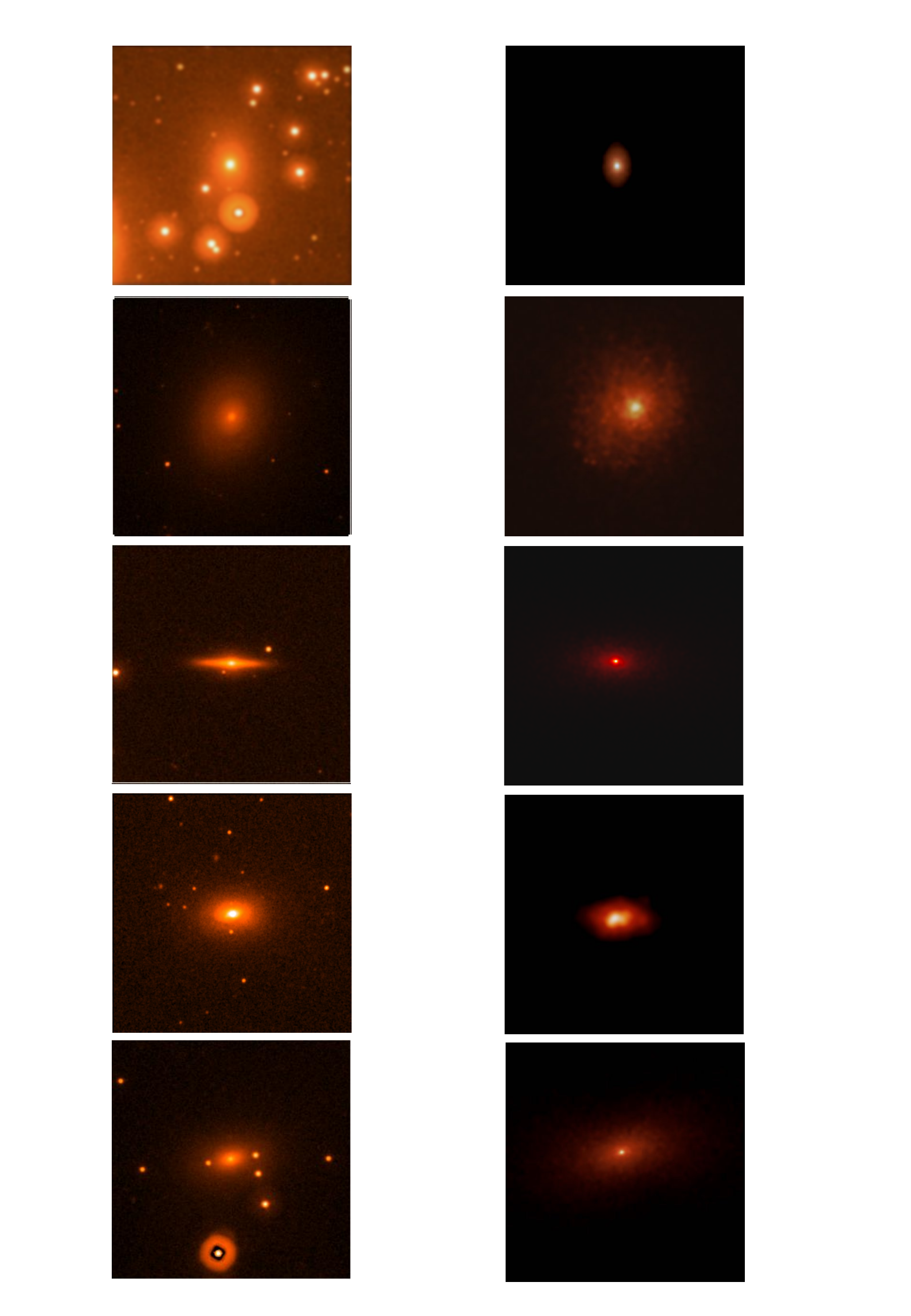}}
 \caption{
 Observed and simulated morphologies of our galaxies in the  UVOT- M2  band (col. 1 and 2) and in the UVOT-V band (col. 3 and 4).
The 
field of view is  5\arcmin$\times$5\arcmin and the resolution 1 pix~arcsec$^{-1}$. 
Simulated morphologies come from simulations in Table~\ref{table6} at the same 
snapshots matching  all the other galaxy properties (Section \ref{simul}).}
\label{fig9}
   \end{figure*}

\begin{figure*}
  \centering
 {\includegraphics[width=11cm]{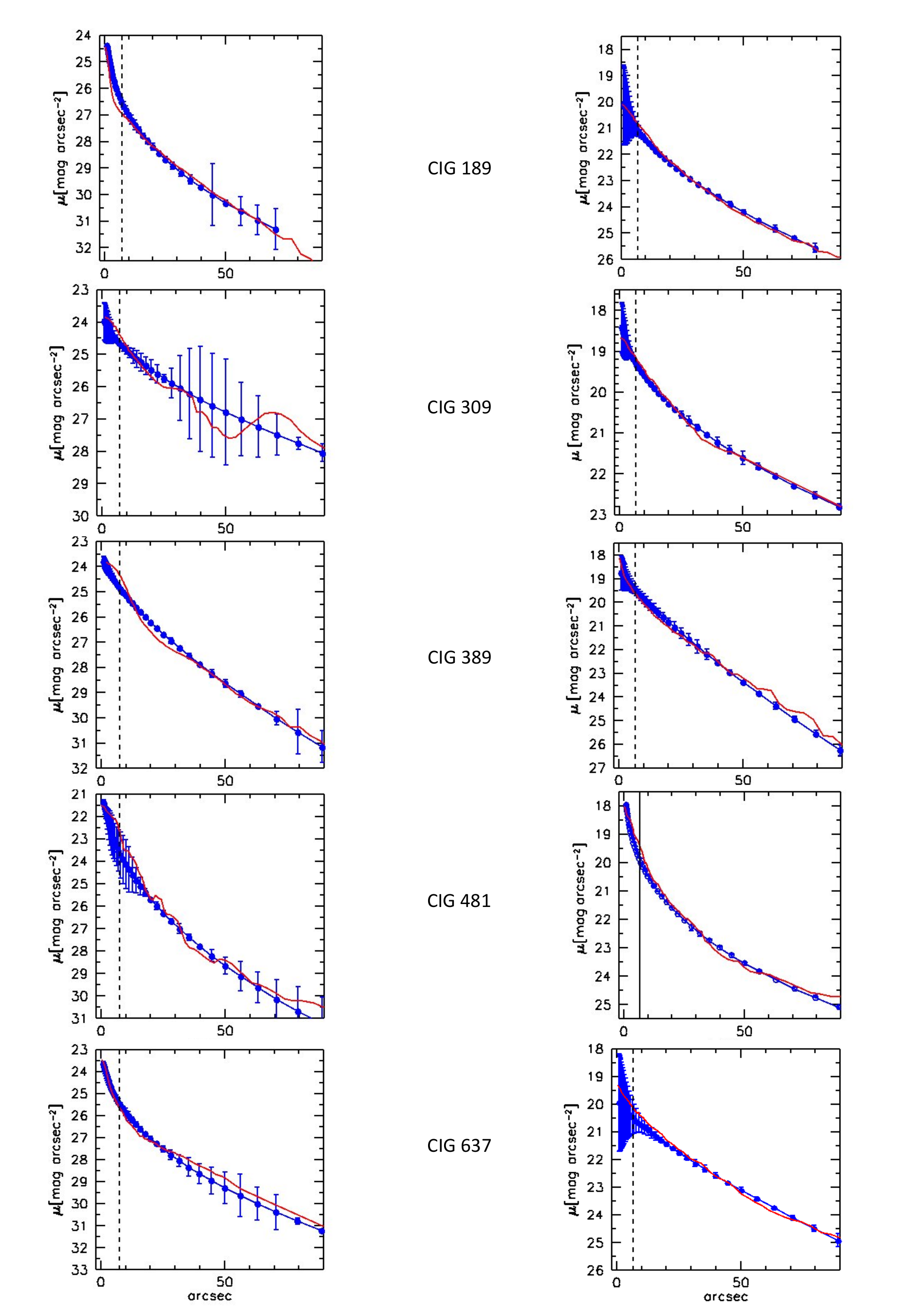}}
      \caption{Luminosities profiles of our targets  in the Swift-UVOT M2
(left) and V (right) bands. Blue lines are best-fit S\'ersic laws in
Section 4.2 and Table \ref{table5} and red  lines are from the simulation snapshots in
Figure \ref{fig9}.
}
         \label{fig10}
   \end{figure*}

We are exploring galaxy formation and evolution via a merger/interaction scenario focusing on low density
environments. We have performed a large grid of simulations of galaxy
encounters and  mergers  starting from collapsing triaxial systems
composed of DM and gas as described in several previous works
\citep[][and references therein]{Mazzei2014a, Mazzei2014b, Mazzei2018, Mazzei2019}.
We combine Smooth Particle Hydrodynamic (SPH) code with
Chemo-Photometric Implementation (SPH-CPI simulations hereafter) based
on Evolutionary Population Synthesis (EPS) models providing the spectral
energy distribution (SED) at each snapshot. The SED accounts for
chemical evolution, stellar emission internal extinction, and
re-emission by dust in a self-consistent way 
 (Spavone et al. 2009, 2012, and references therein). This extends
more than four orders of magnitude in wavelength, from 0.05 to 1000\,$\mu$m. Each simulation in the grid self-consistently provides
morphological, dynamic, and chemo-photometric evolution.

From our  grid of SPH-CPI simulations,  for each galaxy we select the
simulation that matches its current, global properties, in particular
accounts for the following observational constraints: (i) the total absolute B-band magnitude within the range allowed by observations (Table 1), (ii) the integrated SED extended over four orders of of magnitude in $\lambda$,,  (iii) UV and optical morphologies as confirmed by a (iv) comparison between luminosity profiles.
 Moreover, the selected
simulation must account for v) the available kinematic data, and  vi) the HI gas mass in Table \ref{table1}.
  
 These simulations,
anchored {\it a posteriori} to the local properties of our targets, are
used to give insights into galaxy evolution and, in particular, to shed
light on the quenching phase which is identified by the rest-frame NUV-r
vs Mr color magnitude diagram (CMD) that is  an excellent tracer of
even small amounts of residual SF \citep[see e.g.][]{Mazzei2014a}. Therefore, we
analyze here the galaxy transformation by looking at the behavior of  the
SFR, the  mass growth of different components (gas, stars, and DM), the
specific SFR (SSFR) and, in particular, the CMD diagram. This is a
useful tool to capture very low level of residual SFR, highlighting the
length of different phases leading to quenching and galaxy
transformation from the Blue Cloud (BC) to the  Red Sequence (RS). We applied already
the same approach to ETGs belonging to the galaxy groups USGC 376 and LGG 225
\citep{Mazzei2014a}, and to some S0 galaxies, namely NGC 3626 and NGC
1533, in \citep{Mazzei2014b}, where we match their photometric,
structural (e.g. disk vs. bulge) and kinematical (gas vs. stars)
properties, showing that a major merger (1:1) accounts for the
structural and photometric transformations expected in S0s systems
\citep{Querejeta2015}. Furthermore, our approach allowed us to fit the
current global properties of i) several other ETGs
\citep{Spavone2009,Spavone2012,Trinchieri2012, Bettoni2011,
Bettoni2012,Bettoni2014, Mazzei2019}, ii) the blue dwarf galaxy UGC 7639
\citep{Buson2015}, iii) a few late-type galaxies of different luminosity
classes \citep{Bettoni2011, Bettoni2014, Mazzei2018}, iv) a pre-merger
case, NGC 454 \citep{Plana2017}, and v) the false pair NGC 3447/3447A
\citep{Mazzei2018}, showing that this a powerful tool to investigate
galaxy evolution.

Each simulation of our grid of galaxy mergers and encounters starts from  collapsing triaxial halos initially composed of DM and gas \citep{CM98, MC03} built up with the same initial conditions  and the parameters tuned  in previous papers \citep[][and references therein]{Mazzei2014a, Mazzei2014b, Mazzei2018, Mazzei2019}. 
This set accounts for different initial masses (from 10$^{13}$ to 10$^{10}$\,M$_{\odot}$ for each system), mass ratios (from 1:1 to 10:1), gas fractions (from 0.1 to 0.01), and particle resolutions (initial number of gas and DM particles from 6  to 22$\times$10$^4$). By seeking to exploit a wide range of orbital parameters, we varied the orbital initial conditions in order to have the first {\bf pericentre} separation, p, equal to the initial length of the major axis of the more massive triaxial halo down to 1/10 of the same axis for the ideal Keplerian orbit of two points of given masses. For each value of p, we changed the eccentricity in order to have hyperbolic orbits of different energy. The spins of the systems are generally parallel to each other and perpendicular to the orbital (XY) plane. Misaligned spins have also been analyzed in order to investigate the effects of the system initial rotation on the results. Table \ref{table6} reports the initial conditions of each simulation that best reproduces, at a given snapshot, the global current properties of our targets. We recall here that by major mergers, we mean mergers with the initial mass ratio of halo progenitors $\le$4, while in minor mergers, the ratio
 is $>$4.

We refer the interested reader to the paper by \citet[][Section 3 and
references therein]{Mazzei2019} where the recipes of our simulations and
the grid are extensively described. We note that  all the  selected simulations correspond to galaxy mergers, not to encounters (Table \ref{table6}). Moreover,  no single collapsing  halo, like those analyzed in \citet{MC03}, is able to match the global properties of our targets  \citep[][and references therein]{Mazzei2019}.
 The initial gas fraction  is 0.1 in all the cases here analyzed. The stellar mass resolution ranges from 0.4 to 0.04 the gas mass resolution that is, for simulations in Table \ref{table6},   (0.8-20)$\times$10$^{5}$\,M$_{\odot}$. The final number of particles is at least twice that of the initial number in Table \ref{table6}.

\subsection{Results from SPH+CPI simulations}
\label{simul}
From our grid of SPH-CPI simulations \citep[][and references
therein]{Mazzei2019}  we concentrate on the one simulation that provides
a snapshot which convincingly reproduces the global properties of the
target considered, accounting for the observational constraints  at points  i)-vi) of Section \ref{Sims}.
This
snapshot sets the age of the galaxy accounted for from the onset of the 
SF. Table \ref{table7} summarizes the global properties
of the simulations in Table \ref{table6}  from  this snapshot.
The galaxies analyzed
are found to span a large range of ages,   from 8.7\,Gyr for CIG~481,  to
13.7\,Gyr of CIG~189 and CIG ~389, the oldest galaxies in our sample, and
of total stellar masses   1.4 -- 17.6$\times$10$^{11}$ M$_\odot$. The
absolute magnitudes derived from the snapshots best fitting the global
properties of our targets are reported in Table \ref{table7} to be
compared with the observed ones in Table \ref{table1}. Figure \ref{fig8}
allows the comparison between the observed SEDs, extended over almost 4
orders of magnitude in wavelength, and those derived from the selected
snapshot of each simulation (red solid line). Error bars account for
band width (x-axis) and 3$\sigma$ uncertainties of the flux (y-axis).
The snapshot FIR SED is always composed by a warm and a cold dust
component both including PAH molecules as described in
\citet{Mazzei1992}, \citet{Mazzei1994a}, and \citet{Mazzei1994b}. The
warm dust component is heated by massive stars in HII regions, the cold
one by the diffuse light in the galaxy.  More details are given in the
subsections focusing on each galaxy    (Section \ref{sim189}--\ref{sim637}). UV and optical morphologies are
compared with those provided by the selected snapshot in the same bands,
 with the same field of view and resolution  in Figure \ref{fig9},  showing in general a good agreement with global  morphologies. In
order to derive the general trend of the underlying  structure  and to provide a further check for morphologies, the
corresponding luminosity profiles are also compared both in the UV and in the V band (Figure
\ref{fig10}). Properties at points v) to vi) above are discussed for
each target in the dedicated Sections from \ref{sim189} to \ref{sim637}. 
\begin{table}
\centering
\scriptsize
\caption{Properties of five CIG galaxies from simulations}
\begin{tabular}{lccccc}
\hline\hline
Galaxy  & M$_B$           & age       &       M$_{stars}$                    & M$_{tot}$\\
ID         & [mag]              &[Gyr]        &  [10$^{10}$ M$_{\odot}$]   & [10$^{10}$ M$_{\odot}$] \\
         &                       &             &  M$_{R_{25}}$   M$_{R50}$  &       M$_{R_{25}}$   M$_{R50}$    \\
 \hline
CIG~189      &        -18.98    &                13.7                  &    4.67 --   7.50    &  6.91 -- 22.79 \\
CIG~309     &         -20.61     &               10.2          &          12.65 --17.60         &19.68 --55.57 \\
CIG~389      &         -19.25     &               13.7          &         2.50 --  8.47        &      3.45 -- 26.32 \\
CIG~481     &         -19.11    &                  8.7           &          1.97 --   6.23        &     0.75 --  24.91 \\
CIG~637     &         -19.40    &                  9.6           &          3.50 --  7.23        &   8.18 --   9.54 \\
\hline

\end{tabular}
\label{table7}

\tablecomments{ Intrinsic  B-band total absolute magnitude in col.1, the galaxy age in col.2, the 
stellar mass derived within our selected reference radii, R$_{25}$ (Table 1) and R$_{50}$ (50\,kpc) in col.3, 
(left and right respectively), and the corresponding total mass in col.4, (left and right.)}
\bigskip
\end{table}

Relevant evolutionary properties are presented from Figure \ref{fig11} to Figure
\ref{fig15}.  These include four panels for each target highlighting
connections between general evolution and the path of the galaxy in the
rest-frame optical-UV CMD which reflects the behavior of the SFR and
emphasizes how each galaxy transforms and quenches. 
 The evolution is stopped at 14\,Gyr (cf. Section
 \ref{intro}) in all these panels.
The SFR, {\bf in panel a)}, is shown
with a blue solid line that becomes red as the merger phase begins, namely, when
the two stellar systems can no longer be distinguished because their
mass centers cannot be identified/separated anymore. We find that the
SFR shows a gentle self-quenching after reaching its maximum value, due
to gas exhaustion and stellar feedback, lasting several Gyr in all of
our targets.

The path of the galaxy in the
rest-frame  CMD  is shown in panel b) with a blue solid line.
In this panel
  we assume, as \citet[][and references therein]{Mazzei2019}, that the  Green Valley (GV) lies between NUV-r $=$3.5,  which marks the GV
  entry, and NUV-r$=$5, the RS threshold,   (green dashed lines). Significant evolutionary
  stages of each target, that is the brightest point (BP in the following), the entry in the GV, and the RS achievement, are outlined by (black) squares; (red) triangles correspond
  to z=1 and z=0.5 using the cosmological parameters in Section
  \ref{intro}.  The loci of local (z=0) galaxies in the same panel   are reported
  following prescriptions of \citet{Wyder2007} in magenta for the RS and cyan for the BC.  Following our simulations, each galaxy moves along
  the BC  until reaching its maximum SF which corresponds to the
 BP of its rest-frame CMD. Then, the SFR fades,
  and the quenching begins. This causes the crossing of the BC and GV.  Table \ref{tabnewb} reports how long these meaningful evolutionary phases of our targets are.
\begin{table}
\centering
\scriptsize
\caption{Length of several evolutionary steps}
\begin{tabular}{lcccccc}
\hline\hline

Galaxy& $\Delta$t$_{BP}$ & $\Delta$t$_{BP-BC}$ & $\Delta$t$_{GV}$ & $\Delta$t$_{RS}$  \\ 
   ID          & Gyr  &  Gyr  &  Gyr  &   Gyr     \\          
\hline           

CIG~189& 4.14 & 0.80& 1.6& 5.5 \\  
CIG~309&  1.33 & 3.77& 0.91  & --- \\ 
CIG~389&  4.41 & 2.45 & 3.00 & 2.62   \\  
CIG~481&  2.72& 1.62 & 1.60 & ---\\ 
CIG~637 & 3.70&    0.30 & 2.08  & 2.68  \\ 
 \hline
\end{tabular}

\tablecomments{ $\Delta$t$_{BP}$  indicates how many time the simulation needs to  get the Brighter Point  \label{tabnewb} on its rest-frame CMD, $\Delta$t$_{BP-BC}$ the time range to get BC from BP, $\Delta$t$_{GV}$  to cross the GV, and $\Delta$t$_{RS}$ the time spent in the RS.}
\end{table}
  Three galaxies of our sample   belong to the RS, CIG~189, CIG~389, 
  and CIG~637. CIG~309 and CIG~481 are still in the GV. 

 The mass assembly history, that is, the evolution of different mass
 components, is shown  in panel c) within a fixed radius of 50\,kpc centered on the
 B-band luminous center of the galaxy. The black solid line is the total
 mass, the red short-dashed line is the DM, the blue dotted line is the
 gas +stars (baryons), the magenta long-dashed line is the stellar
 component, and the green  dot-dashed line shows the amount of gas with
 temperature lower than 2$\times$10$^4$\,K.  Here, as in the following,
 we refer to  this gas component as to cold gas given that   its cooling
 timescale is shorter than the snapshot time resolution (0.037\,Gyr,
 \citet{Mazzei2019}).  This represents the upper limit of H I provided
 by each simulation.  
 The percentages of stellar mass accreted  in relevant evolutionary redshift intervals  are collected in Table \ref{tabnew}.
  Panel d) shows the behavior of the SSFR which further highlights the galaxy transformation during  evolution.
  The dashed horizontal line in this panel corresponds to the threshold value between star forming and quenched galaxies, 3$\times$10$^{-10}$ yr$^{-1}$ \citep{Eales2017}.
  
 Figure \ref{fig18} and Figure \ref{fig19} collect projected morphologies  at redshift 0.24, that is 3\, Gyr ago compared to the current
age of each target galaxy, and at  BP of their rest-frame CMD respectively, as observed in the V-band and with a field of view of 20\arcsec.
CIG~309 and CIG~481 have distorted morphologies, very different from the current ones, and CIG~389  and CIG~637 appear as barred galaxies at redshift 0.24. Only  morphology of CIG~189 agrees with its current one, as expected for an isolated  evolution. 
Moreover, to conclude our evolutionary picture,
Figure \ref{fig19} highlights that complex morphologies appear as the galaxy achieves its  BP of the rest-frame CMD, 
showing  tails and features like jelly-fish galaxies \citep{Roman2019}. 
We direct the interested reader in  the evolution of our targets to the following, devoted,  subsections (Section \ref{sim189} --\ref{sim637}) providing further details 
 for each target and  related Figures (Fig.\ref{fig8} --Fig.\ref{fig19}).
\begin{table*}
\centering
\scriptsize
\caption{Percentages of the stellar mass growth in relevant evolutionary redshift intervals}
\begin{tabular}{lcccccccc}
\hline\hline

Galaxy&z${_i}$&$\Delta$z(${_i}$--0) & $\Delta$z(${_i}$-2) & $\Delta$z(1--0) & $\Delta$z(0.5--0) & $\Delta$z(BP--0) & $\Delta$z(GV) & $\Delta$z(0.24--0) \\ 
   ID       &   &  {\%}  &{\%}   & {\%}  & {\%}     & {\%} &{\%}   &  {\%}  \\          
\hline           

CIG~189&3.45& 94.9 & 24.2& 9.1& 0.12 &  5.1  &2.8& 0.5 \\  
CIG~309& 0.62& 60.0 & ---&  85.2 & 47.8 &37.8  &1.1& 17.8\\  
CIG~389& 4.52& 98.5 & 13.5 & 33.1 & 4.3  &35.1 & 1.7  &1.2 \\  
CIG~481&0.60&  83.0& --- & 99.9 &91.2  &17.5& 2.6 & 16.0 \\  
CIG~637&1.20 & 97.7& ---&  87.0 & 7.15& 1.6  &-0.07  & 1.6\phantom{0}  \\ 
 \hline
\end{tabular}

\tablecomments{ In col.2 z${_i}$ indicates  redshift of merger beginning as defined in Section 5.1 with   cosmological parameters in Section
\ref{intro}. In col.s 3-- 7: percentages of stellar mass within  50\,kpc;  BP is the Brighter Point  and GV the  Green Valley of the CMD of each galaxy.  NB:  negative fractions indicate mass loss in the interval considered. }
\label{tabnew}
\end{table*}
\subsection{CIG 189}
\label{sim189}

The simulation that best matches the global properties of this ETG
corresponds to a minor merger (5:1) between two halos, each of gas and DM,
with perpendicular spins at the beginning; the other input parameters
are in Table~\ref{table6}. The merger occurs  1.7\,Gyr  after the onset
of the SF, at redshift 3.45  (Table \ref{tabnew}).  At this age the two galaxies begin to mix and, 1.2\,Gyr later,
at redshift 2.2, the  galaxy appears  with a long tail since the
instability due to the merger is  still  working. From the snapshot that
best reproduces its global properties  (Figure~\ref{fig8},
Figure~\ref{fig9} and Figure~\ref{fig10})  this galaxy  is 13.7\,Gyr old with
a B-band absolute magnitude of about -19\,mag (Table \ref{table7}), in  agreement with the value in
Table~\ref{table1}. The age of the galaxy derived from the mean age of
its stellar populations within 50\,kpc is 10.7\,Gyr  and 9.6\,Gyr
weighting by their B-band luminosities. The same quantities computed
inside a radius equal to R$_{25}$ (Table \ref{table1})  become 9.9 e 
9.2\,Gyr. Both these estimates point towards a red and dead galaxy.
Looking at its SED in Figure~\ref{fig8}, where  cyan square in the
far-UV is from the {\tt GALEX}  archive, we note that the FIR SED is as expected for
ETGs  \citep{Mazzei1994a, Mazzei1994b}. However its  warm-to-cold dust
energy ratio  is almost ten times lower than expected,   0.02,
due to the constraint given by  the WISE datum  at 11\,$\mu$m. The
central (r$\leq$1 kpc) velocity dispersion at the same snapshot, 164.9\,
km\,s$^{-1}$, agrees with the value in the HyperLeda catalog,
147.8$\pm$13.6\,km\,s$^{-1}$. There are no other available global
observational constraints.
\begin{figure*}
  \centering
 {\includegraphics[width=15.6cm]{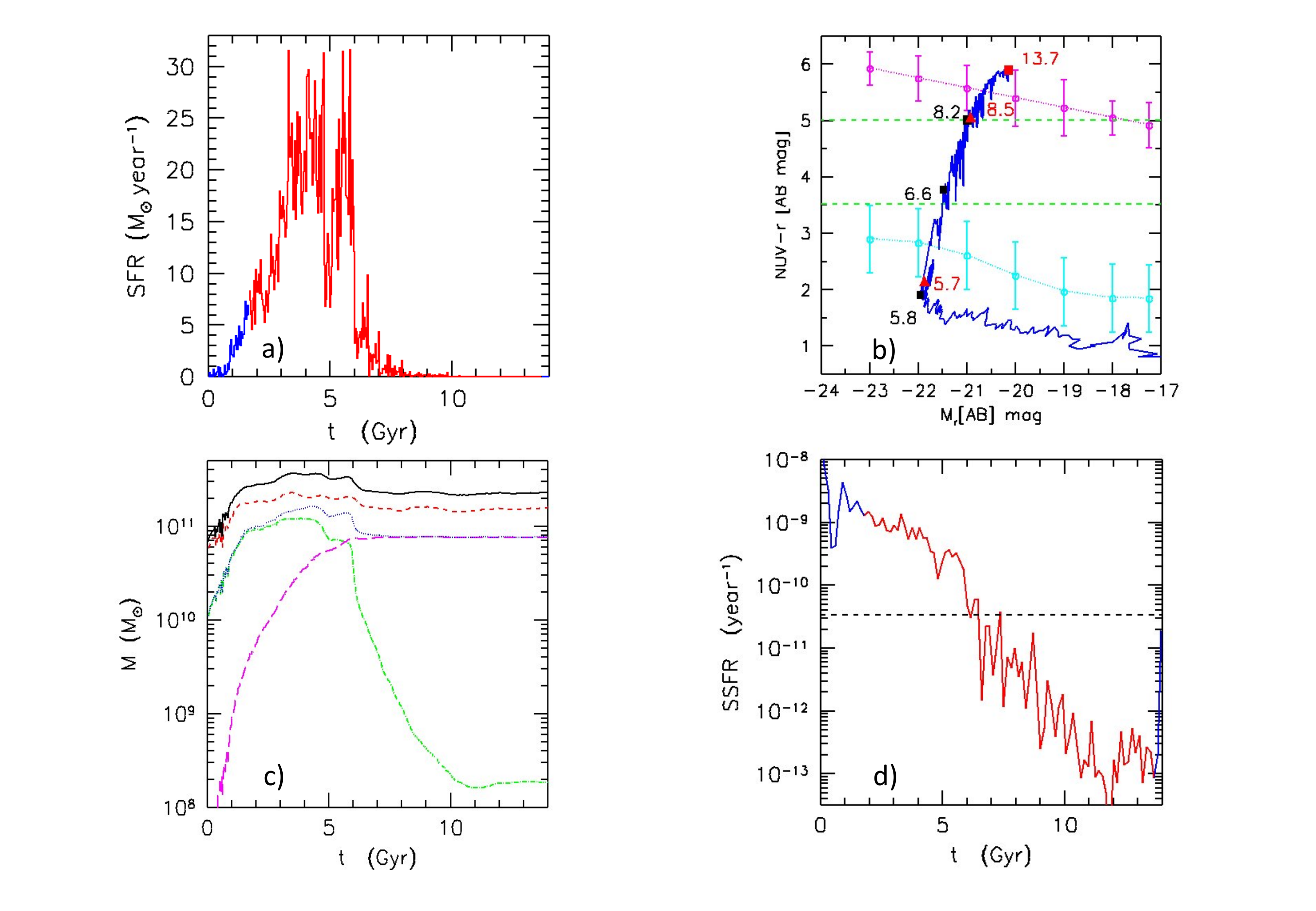}} \caption{Evolutionary
properties of CIG~189 within 50\,kpc radius in the B-band  (Section \ref{sim189}).  (a) The SFR, 
red  from the beginning of the merger  (as defined in Section \ref{simul}) to the current galaxy age (Table \ref{table7}).  (b)
The evolution in the CMD of   galaxy rest-frame  (blue solid line);  some important galaxy ages, including
its current age  (red filled square), are also reported (more details in Section  \ref{simul}).  (c)
The growth of the mass components: stars (magenta, long-dashed), DM (red, short dashed), 
baryons (blue, pointed)), gas with T$<$20000 K, that is the upper limit of cold
gas (green,  dot-dashed),  and the total mass (black, solid line).  (d) The SSFR, red from the beginning of 
the merger to the current galaxy age; the dashed horizontal line corresponds to the threshold
value between star forming and quenched galaxies \citep{Eales2017}.   Section \ref{sim189} for more details.
}
       \label{fig11}
   \end{figure*}
Looking at its evolution (Figure~\ref{fig11}),  this galaxy crossed the
BC in about 1.9\,Gyr, the GV in 1.6\,Gyr, then  lived 5.5\,Gyr on the RS
reached at redshift 0.54 (Table \ref{tabnewb} and panel b). The evolution we derived agrees well with
the idea  of an isolated red and dead ETG, as suggested by the previous
analysis of its  local (z=0) photometric (n$>$2.5, Section \ref{surfphot}) and kinematical properties. Moreover, its  SSFR has fallen
below  the critical value of star forming galaxies,
3$\times$10$^{-10}$ yr$^{-1}$ \citep{Eales2017} from an age of about 6
Gyr (panel d),  corresponding to the BP on its CMD, and
it has remained below this threshold until its current age.
This behavior  has covered a time range of  7.7\,Gyr.
Therefore,  at redshift 0.24, that is 3\, Gyr ago compared to its current
age,  the galaxy lies on the RS with a SSFR lower
than 1$\times$10$^{-12}$\, yr$^{-1}$, and the
morphology   expected for an ETG (Figure \ref{fig18}).

Panel c) of Figure \ref{fig18} and Table \ref{tabnew}  highlight that  i) about 95\%  of the stellar mass was assembled from the beginning of the merger  to the current galaxy age,  ii)  71\% from redshift 2,  and iii)  90.9\% (78.4\%) of the current stellar (total)  mass was in place at redshift 1. 
iv) Only 2.8\% of its current stellar mass has been assembled during the crossing of the GV and, v) 0.5\% in the last 3\,Gyr, that is from redshift 0.24 to z$=$0.

We point out  that  V-band observed morphologies  of this galaxy at its BP are like a jelly-fish galaxy (Figure \ref{fig19}) as defined by \citet{Roman2019}.

\subsection{CIG 309}
\label{sim309}
The simulation which best matches the global properties of this galaxy
(from Figure~\ref{fig8} to \ref{fig10})
corresponds to a major merger of three halos occurred  6\,Gyr in the past of the galaxy, at redshift 0.62.
The galaxy is 10.2\,Gyr old
(Table~\ref{table7}). Its   total B-band magnitude 
(Table~\ref{table7}), agrees, within the error,   with the value in
Table~\ref{table1} (see also Appendix). Its SED (Figure~\ref{fig8})  includes {\tt GALEX} data by
\citet{Bianchi2017},  violet squares,  and  FIR data, yellow dots, 
from the AKARI catalog. The   FIR SED is well fitted by an early-type
dust-stars distribution, as in \citet{Mazzei1994a}. However, 30\% of its
bolometric luminosity   comes out in the FIR spectral range, ten times
more than expected on  average for ETGs,  like a normal Spiral
\citep{Mazzei1992}. The age of the galaxy derived from the mean age of
its stellar populations within 50\,kpc is  7.8\,Gyr  and 7.2\,Gyr
weighting by its B-band luminosities. The same quantities computed
within R$_{25}$ in Table \ref{table1}, corresponding to about 10\,kpc,
are 5.7 and 3.5\,Gyr respectively.
The central (r$\le$1\,kpc) velocity dispersion, 174.6\,km\,s$^{-1}$, and
the maximum gas rotation velocity, 288\,km\,s$^{-1}$, at the same
snapshot, agree well with kinematical data reported in Appendix and Section
\ref{surfphot}  showing the values of  S\'ersic indices all below 2.5 (Table \ref{table5}). 
Therefore, from
our multi-wavelength analysis of its luminosity
profiles (Section \ref{surfphot}), matched well by  the selected snapshot (Figure \ref{fig10}), and total SED  accounting for  30\%
of  the bolometric luminosity in the FIR spectral range,  we conclude for a disk structure of this galaxy.
\begin{figure*}
  \centering
 {\includegraphics[width=15.6cm]{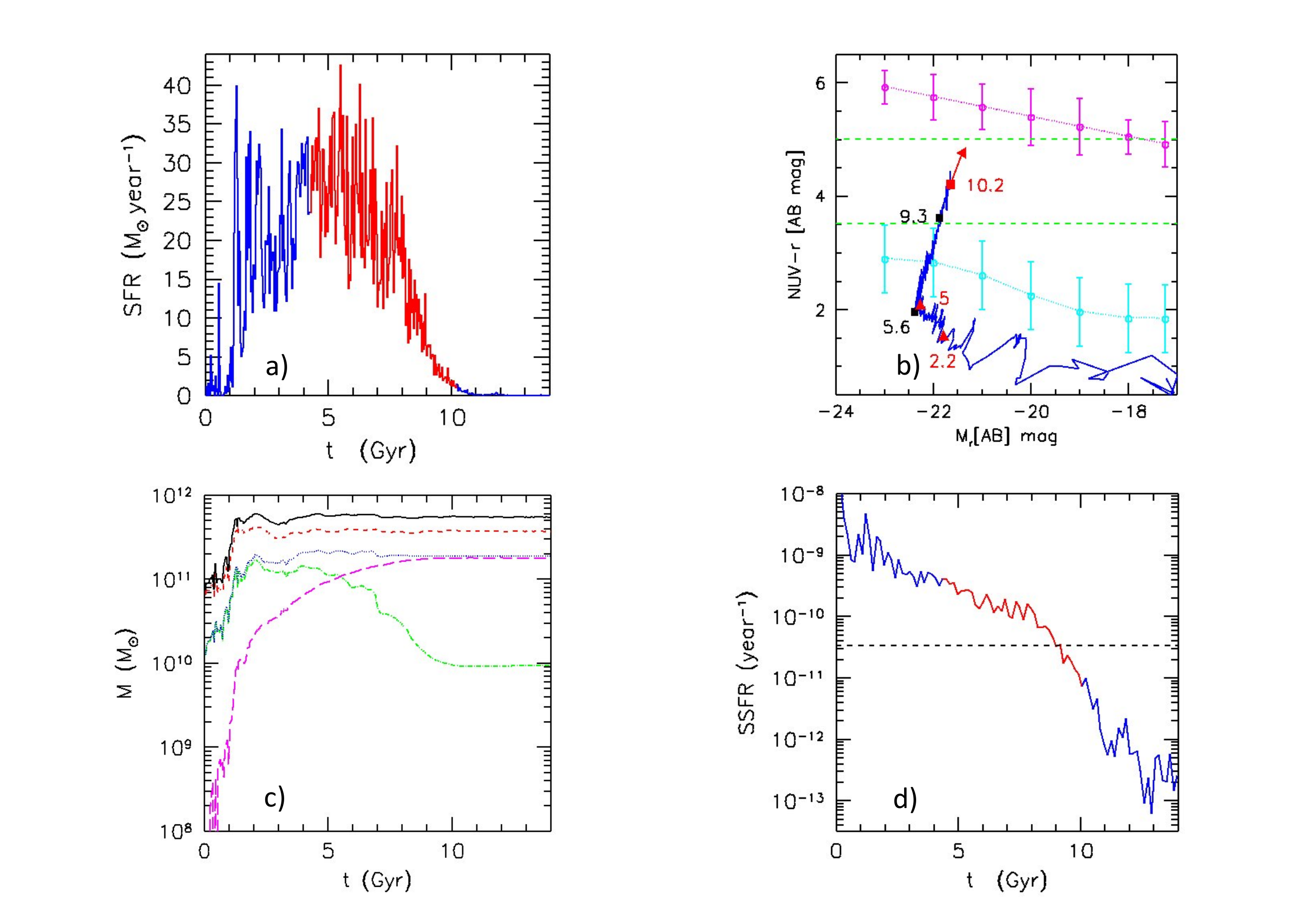}}
      \caption{As in Figure~\ref{fig11} for CIG~309;  Section \ref{sim309} for more details.
}
       \label{fig12}
   \end{figure*}

\begin{figure*}
  \centering
 {\includegraphics[width=15.6cm]{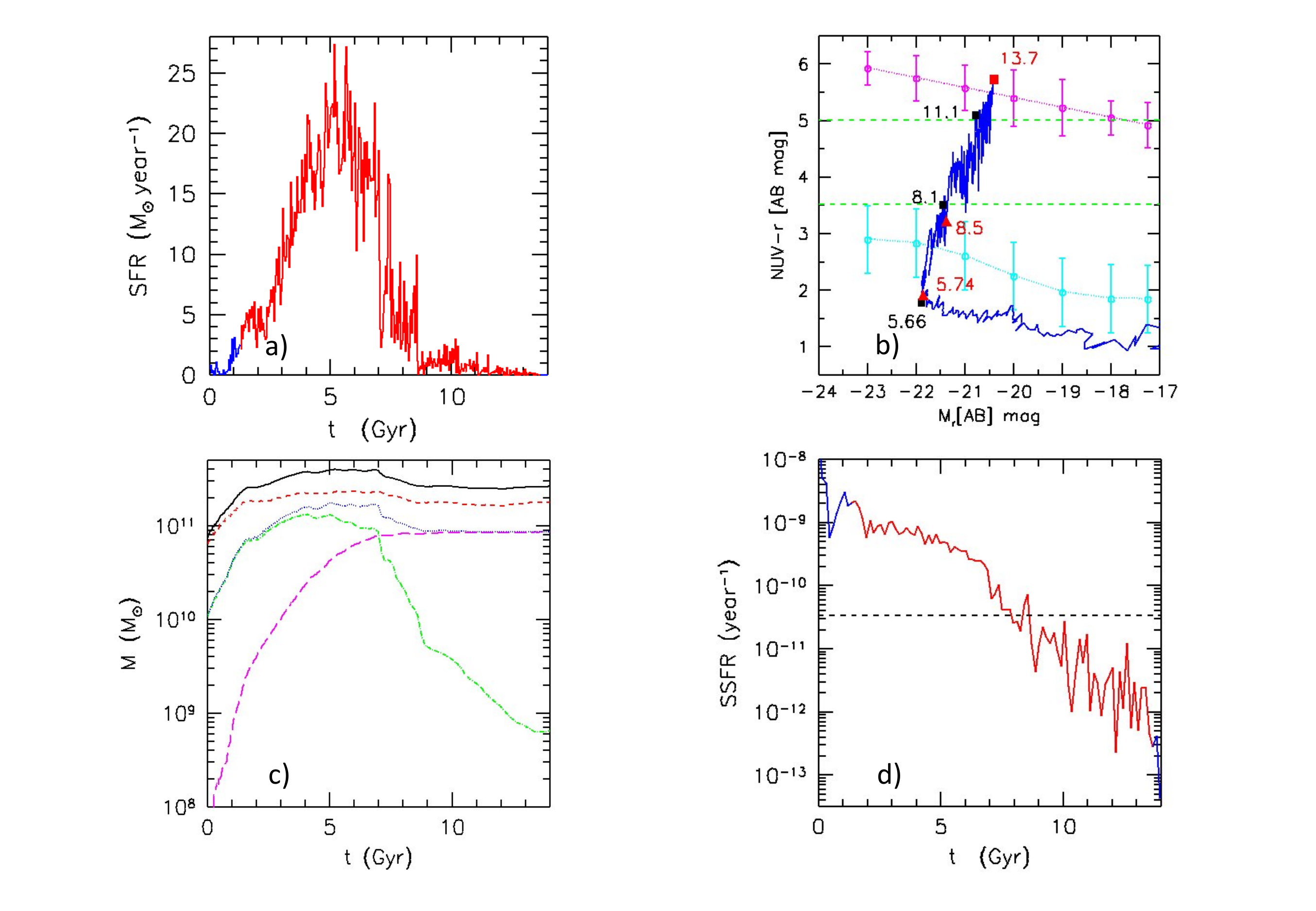}}
      \caption{As in Figure~\ref{fig11} for CIG~389;  Section \ref{sim389} for more details.
}
       \label{fig13}
   \end{figure*}
Looking at its evolution,  the
first stars appear in a  filament  about 500\,kpc long.  A region of
this filament  collapses forming a  more dense body in about 1\,Gyr from
the onset of the SF.  This galaxy  is connected to a small mass
companion, like a plume,  from one side,  and a tail from
the other side 0.3\,Gyr later. The final merger,  that is  a unique,  quite irregular,
stellar system  with a plume or shell that appears as a long tail  in edge-on
projections, was formed  6\,Gyr  ago, at redshift 0.62  according our
cosmology (Section 1).  The galaxy lies in the BC at that time, looking at
its rest-frame CMD, with a SSFR larger than the threshold value, that is
like a late-type galaxy (Figure~\ref{fig12}, panels b and d).  The galaxy leaves  the BC 3.8\,Gyr later   (Table \ref{tabnewb}), at an age
of 9.4\,Gyr and reaches its current position on the CMD, with a SSFR  
below the threshold value,   in less than 1\,Gyr. 
At redshift 0.24, that is 3\,Gyr before its current age,  at 7.2\,Gyr, 
this galaxy was on the BC with a SSFR like a spiral system. 
 Its morphology  (Figure \ref{fig18}), is quite disturbed showing an off-centre nucleus
and a long plume in x-y projection.  All its properties are very different from the current ones.

From Figure \ref{fig12}  (panel c),  we derive that
i)  15\%  of the current stellar  mass was in place at redshift 1. Table \ref{tabnew}  highlights that ii) about 60\%  (2.2\%) of the stellar (total) mass was assembled from the beginning of the merger, z=0.62, to its current  age, iii)  47.7\%  of the current stellar  mass is build up from redshift 0.5 to 0.
However, in the same range, the total mass grows  by only few percent (1.2\%)  while the mass of the gas is about 10 times its current value.
iv) 17.8\% of its stellar mass has been assembled in the last 3\,Gyr. v) The galaxy is still crossing the GV and the stellar mass assembled in this phase is 1.1\% of its current mass. We note that complex
morphologies arise as the galaxy   approaches its highest values of the SFR (Figure \ref{fig19}),
showing tails and features like jelly-fish galaxies \citep{Roman2019}.

\subsection{CIG 389}
\label{sim389}

The simulation which best matches the global properties of this galaxy corresponds to a 
minor merger (10:1) with  a  total initial mass of  2.2$\times$10$^{12}$ M$_\odot$  
(Table~\ref{table6}).  

The merger between the stellar systems occurs  1.25\,Gyr  after the
onset of the SF, that is at redshift 4.5  (Table \ref{tabnew}). The snapshot that best reproduces the
global properties of this galaxy (Figures~\ref{fig8}, 8 and 9) suggests
this is 13.7 Gyr old with a B-band absolute magnitude, -19.25\,mag (Table
\ref{table7}), in  agreement with the value in Table~\ref{table1}. The
age of the galaxy derived from the mean age of its stellar populations
within 50\,kpc is  8.7\,Gyr  and 8.3\,Gyr weighting by their B-band
luminosities. The same quantities become  7.4 e 6.8\,Gyr respectively
computed within R$_{25}$. Looking at its SED  in  Figure \ref{fig8}, the
azure square in far-UV  is provided by the {\tt GALEX} archive, and  the 
yellow upper limits are from AKARI catalog. The far-IR SED, well matched
  by the FIR distribution of \citet{Mazzei1994a},   suggests a FIR to
bolometric luminosity ratio of 0.07, as expected on average  for ETGs
\citep{Mazzei1994b}. From the selected snapshot we derive a central
(r$\le$1\,kpc) velocity dispersion of 109\,km\,s$^{-1}$, a maximum
stellar velocity of  135\,km\,s$^{-1}$, and maximum velocity rotation of
cold gas, that is the gas with temperature lower than 20000\,K, of
127\,km\,s$^{-1}$, in agreement with the values in Appendix.
Figure \ref{fig10} shows that the UV and optical profiles of the selected simulation
match well the observed ones which, from the results in Setcion 
\ref{surfphot},  correspond to  S\'ersic indices all well below 2 (Table \ref{table5}).

 Looking  at the evolution in the rest-frame CMD  (Figure~\ref{fig13},  panel b)
 this galaxy moved from its BP at redshift 1.03, when its age was
 5.7\,Gyr,  then  crossed the BC in the following  2.4\,Gyr,  and the
 GV in 3\,Gyr reaching the RS at redshift 0.21. The  galaxy got its
 current position on the  RS  2.6\, Gyr later   (Table \ref{tabnewb}), at the age of 13.7\,Gyr.
 The quenching phase, starting at the beginning of the GV (BC
 point),   is  further  emphasized   by the behavior of the SSFR (Figure~\ref{fig13}, panel d) that 
 becomes lower than the threshold value for ETGs,  3$\times$10$^{-10}$
 yr$^{-1}$ \citep{Eales2017},  just from the age of about 8\, Gyr when
 the galaxy comes close to the entry in the GV.
At redshift 0.24, that is 3\,Gyr ago, its B-band magnitude
was  0.5\,mag brighter  than its current value and the galaxy was  still crossing  the GV.
 Its morphology was  that of a barred galaxy  (Figure \ref{fig18}).

Figure \ref{fig13}  (panel c) and Table \ref{tabnew} highlight that  i) about 98\%  of the stellar mass was assembled from the beginning of the merger to the current galaxy age,  ii) 87.8\% (26.2\% of the current total mass) from redshift 2, and iii)  33.1\% (4.9\%) from redshift 1 to the current galaxy age.
iv) Only 1.7\% of the stellar mass has been assembled during the crossing of the GV and 1.2\%
in the last 3\,Gyr, that is from redshift 0.24 to z$=$0. 

We point out that  for this galaxy too, complex
morphologies arise at the BP point  (Figure \ref{fig19}),  
such as tails and features like jelly-fish galaxies \citep{Roman2019}.

\subsection{CIG 481}
\label{sim481}

\begin{figure*}
  \centering
 {\includegraphics[width=15.6cm]{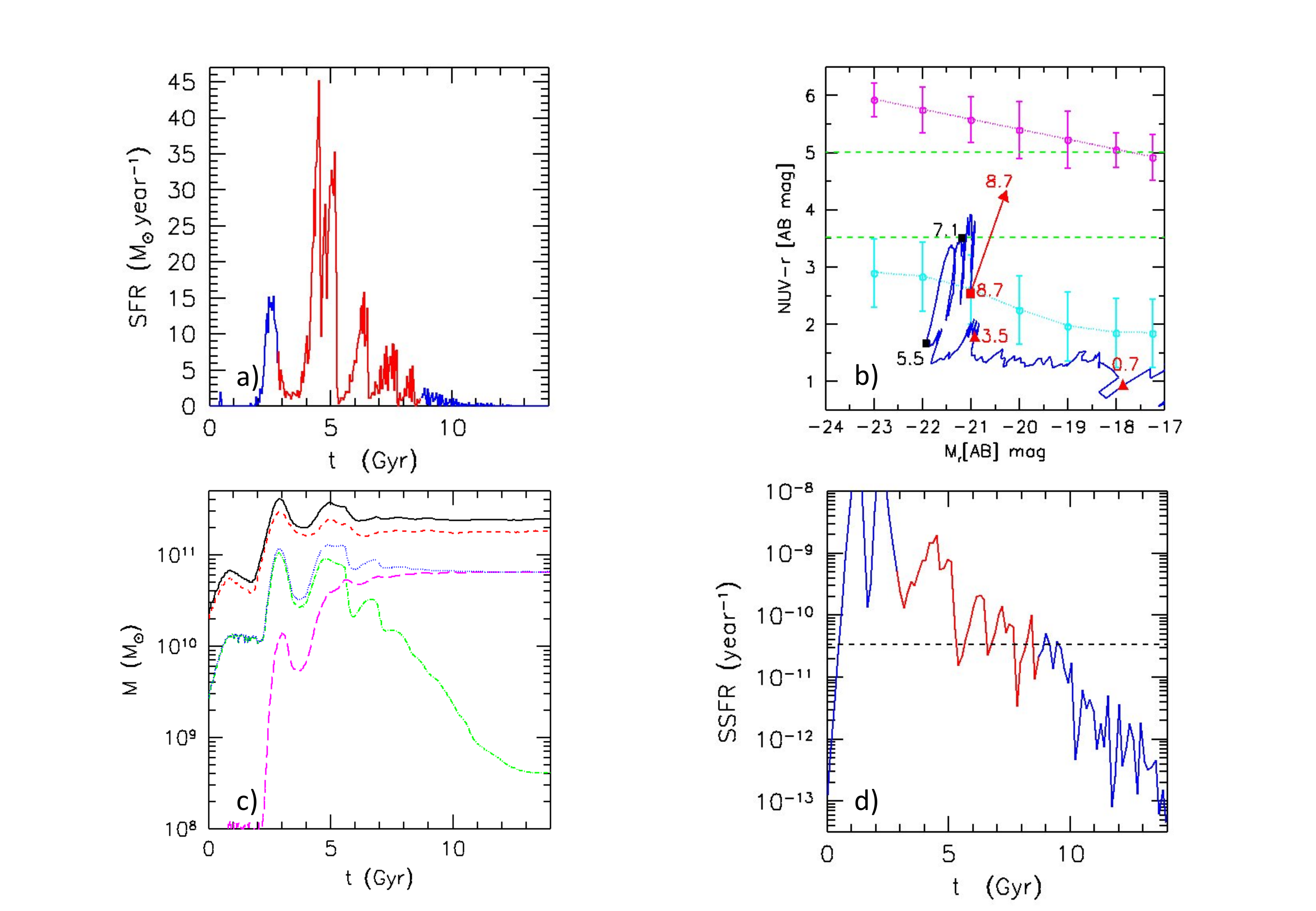}}
      \caption{As in Figure~\ref{fig11} for CIG~481; {\bf Section \ref{sim481} for more details}.
}
       \label{fig14}
   \end{figure*}

The simulation that best matches the global properties of this galaxy
corresponds to a major merger (Table~\ref{table6})  occurred at redshift 0.6, when
the system was 2.8\,Gyr old from the beginning of the SF. The current age
of the galaxy is 8.7\,Gyr , the youngest we derived
for isolated galaxies here, with M$_B$=-19.11\,mag
(Table~\ref{table7}, to be compared with Table~\ref{table1}). The age of
the galaxy derived from the mean age of its stellar populations within
50\,kpc is  5.3\,Gyr  and 2.3\,Gyr weighting by  B-band luminosities;
these values become 3.8 and 1\,Gyr respectively within R$_{25}$ in Table
\ref{table1}.
The total SED  is compared with the observations in Figure~\ref{fig8} where violet squares   are {\tt GALEX}
fluxes from \citet{Bianchi2017}, azure dot and upper limits are from
\citet{Rifatto1995}, yellow points and 60\,$\mu$m upper limit are from
AKAR-FIS  catalog. This galaxy shows a strong FIR emission compared to
normal ETGs. Its FIR SED  requires a warm dust component with a dust
temperature lower than   average for ETGs, 48\,K instead of 63\,K,  but
with a warm-to-cold  energy ratio of 0.7. The bolometric luminosity,
indeed, provides  78\% of the  intrinsic luminosity.  
 There is a large amount of   residual gas 50\,kpc
around the galaxy (Figure \ref{fig14},  panel c) at the  selected  age, about 
4$\times$10$^{8}$\,M$_{\odot}$,  well in agreement with constraint in  Table \ref{table1}.  The maximum  gas rotation velocity,
158\,km s$^{-1}$, is also in agreement with constraint in Appendix.

Looking at the  evolution of the rest-frame CMD (Figure \ref{fig14}  panel b, and Table \ref{tabnewb}),  the galaxy  leaves the
BC at an age of 7.1\,Gyr,  or 1.6\,Gyr ago, at redshift 0.12.  Some
rejuvenation  episodes occur during the crossing of the GV that would make 
its intrinsic color   consistent with the BC.
However, because of internal reddening,  the galaxy remains in the  GV, as  observed.  
 Looking at 3\,Gyr ago,  the quenching phase is 
just begun. The  SF achieved during this time lapse puts the galaxy up and
down  the SSFR threshold value separating late-type from ETGs (Figure
\ref{fig14},  panel d). Residual bursts account for the large shells detected in the galaxy
morphology \citep{Morales18}, which, however, contradicts the strong degree of isolation
established by Verley et al. (2007b, Section \ref{isol}) for this young and
reddened galaxy. These features are the product of the in-situ
evolution, in particular of the evolution of its potential well which is
born and still accreting by the initial merger conditions. Figure
\ref{fig18}  highlights the observed optical morphology of this galaxy
at redshift 0.24, showing that this is quite different from that of a
ETG.

Figure \ref{fig14} (panel c) and Table \ref{tabnew} highlight that  i) 83\% of  its current stellar mass was assembled from 
the beginning of the merger,   ii)  17.5\%  from  the BP,  at redshift 0.27,  to the current galaxy age.
iii) A relatively large amount of its current stellar mass, that is 16\%, has been assembled in the last 3\,Gyr,  and only
v) 2.6\% in the GV.
Figure \ref{fig19} highlights distorted morphologies   at its BP (z=0.27)  as observed in the V-band.

\subsection{CIG~637}
\label{sim637}

This simulation is a merger of three systems with total mass
25$\times$10$^{11}$\,M$_\odot$ (Table 1). The merger occurred after
0.9\,Gyr from the onset of the SF,   at redshift 1.2 (Table \ref{tabnew}). The same snapshot which best
reproduces the properties in Figure~\ref{fig8},  Figure~\ref{fig9}, and 
Figure~\ref{fig10} corresponds to the age  of  9.6\,Gyr and a total B-band
absolute  magnitude  -19.4\,mag  (Table~\ref{table7} and
Table~\ref{table1}). The age of the galaxy derived from the mean age of
its stellar populations within 50\,kpc is  7.3\,Gyr  and 6.5\,Gyr
weighting by   B-band luminosity. The same quantities computed
within R$_{25}$ (Table \ref{table1}) are 6.7 e 6.1\,Gyr.

\begin{figure*}
  \centering
 {\includegraphics[width=15.6cm]{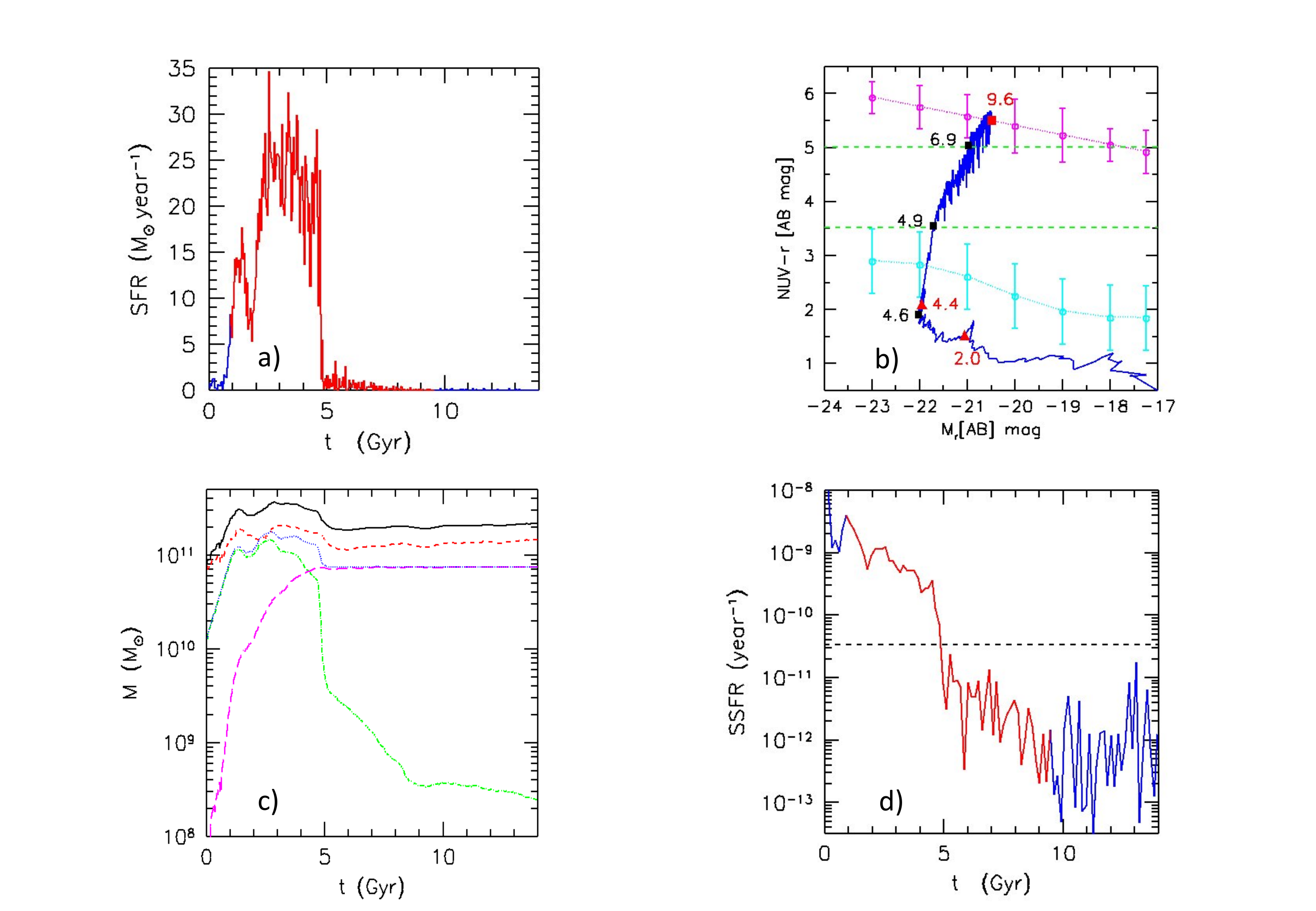}}
      \caption{As in Figure~\ref{fig11} for CIG~637  {\bf (Section \ref{sim637})}.
}
       \label{fig15}
   \end{figure*}

Concerning its SED  in Figure \ref{fig8}, the yellow  upper limits in
the  FIR spectral range are  AKARI data from \citet{Kokusho17}. The  FIR
luminosity is equal to 2\% of the bolometric luminosity, in agreement
with the value expected, on average, for ETGs
\citep{Mazzei1994a,Mazzei1994b}. The amount of cold gas within 50\,kpc
at the age of this galaxy is  2-3$\times$10$^{8}$\,M$_{\odot}$ (Fig.
\ref{fig15},  panel c),  consistent with observations.

  Figure~\ref{fig15} shows the  evolution.

\begin{figure*}
  \centering
 {\includegraphics[width=16.5cm]{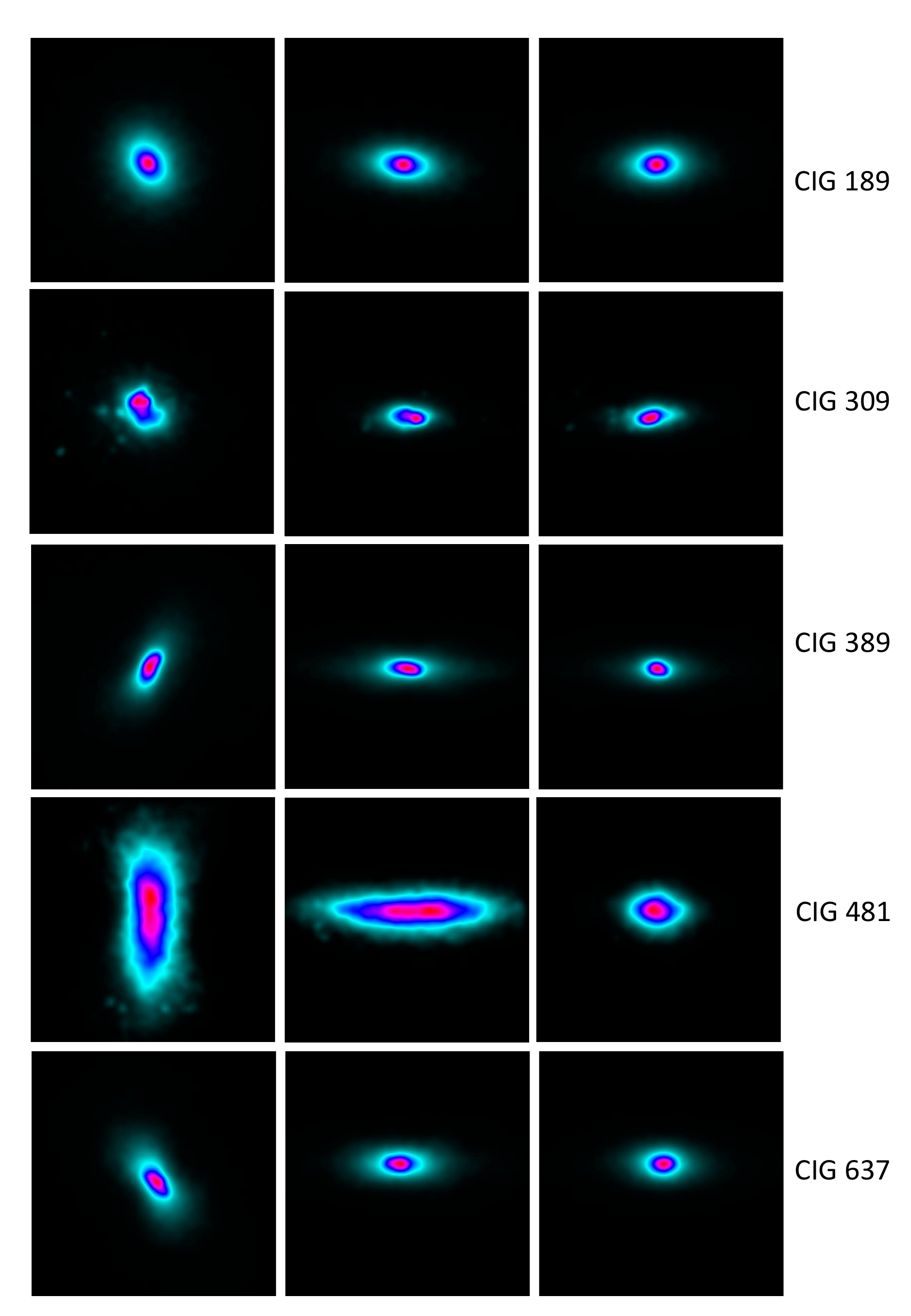}}
 {\includegraphics[width=3cm]{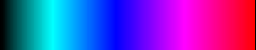}}
\caption{V-band observed morphologies of our targets from simulations in Table \ref{table6} at z=0.24, that is 3\,Gyr before their current age. From left to right: X-Y, Y-Z and X-Z projections, each normalized to the total flux in the frame.  The field of view is 20\arcsec and the resolution  1\,\arcsec\,px$^{-1}$, that is 4.07\,kpc  per \arcsec (Section \ref{intro}).
Bottom: the color scale used. 
}
   \label{fig18}
   \end{figure*}

\begin{figure*}
  \centering
 {\includegraphics[width=16.5cm]{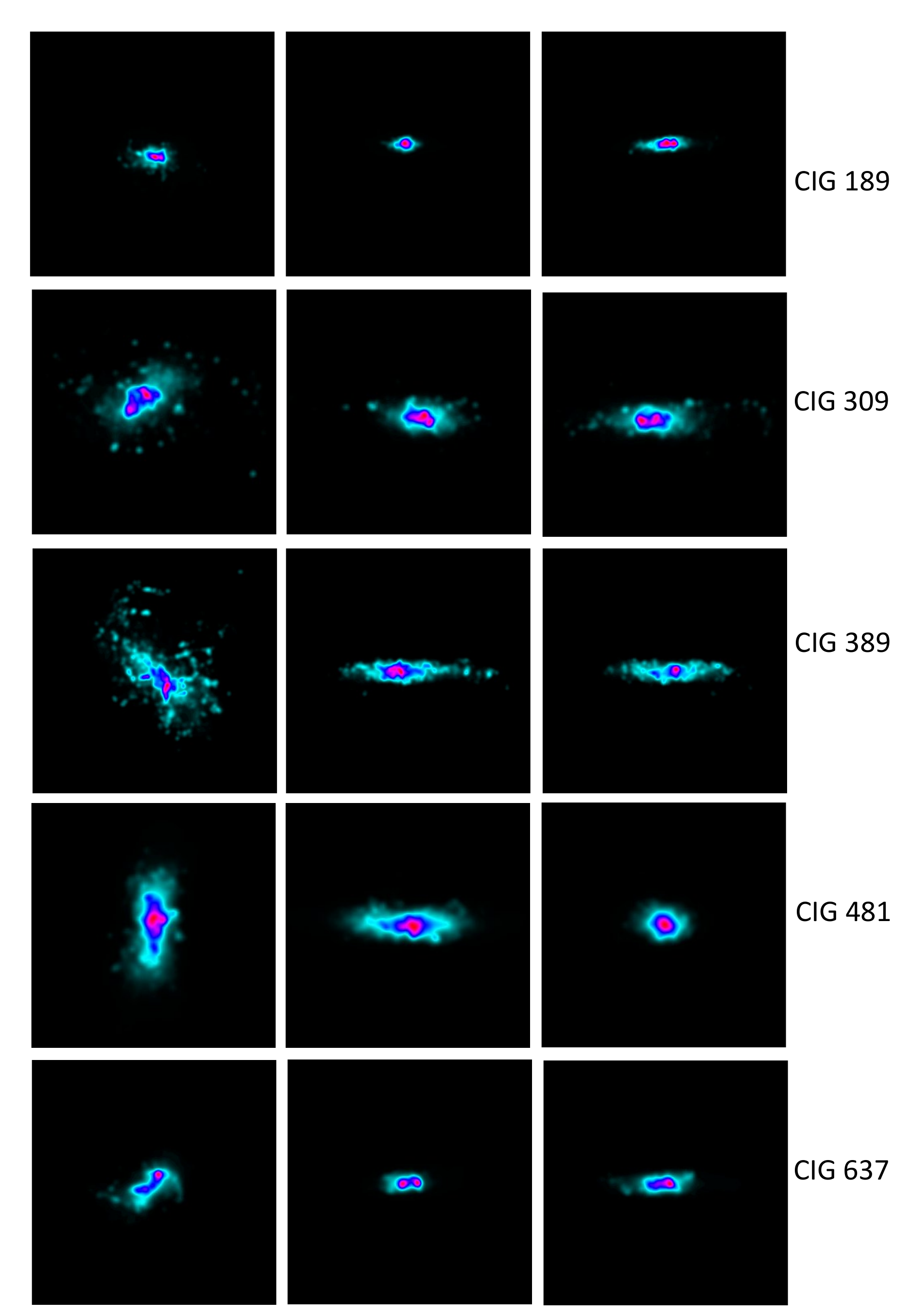}}
{\includegraphics[width=3cm]{Figure_15b.png}}
\caption{Morphologies of our targets from selected simulations as the galaxy lies on the BP of its CMD,  as observed in the V-band. From left to right: X-Y, Y-Z and X-Z projections, each normalized to the total flux in the frame.  The field of view is 20\arcsec$\times$20\arcsec and the resolution 1\arcsec\,px$^{-1}$  (Section \ref{intro}).  Bottom: the color scale used.
}
   \label{fig19}
   \end{figure*}

 The galaxy achieves the  BP on its rest-frame CMD at redshift 0.47,  at an age of
 4.6\,Gyr (panel b). In the next 0.3\,Gyr the galaxy crosses the BC and comes in the GV (4.9\,
 Gyr old).  At the same time its SSFR drops below the critical value
that separates  LTGs from ETGs (panel d).  Following our simulation,  this ETG
 lived in the RS by about 2.7\,Gyr, crossing the GV in the previous 
 2\,Gyr (Table \ref{tabnewb}, 8.7\,Gyr old). Thus,  3\,Gyr ago, at redshift 0.24, this galaxy
 was still crossing the GV showing a shell morphology (Figure \ref{fig18}, bottom). Its SSFR was below the threshold
all this time.
From Figure \ref{fig15} (panel c)  and Table \ref{tabnew} we derive  that i) 97.7\% of the current stellar mass was assembled from the beginning of the merger
 to the current galaxy age,
ii)  87\%   from redshift 1,   iii)  1.6\%  from  the BP, at redshift 0.46,  and almost the same amount  in the last 3\,Gyr.
Projected morphologies of this galaxy as observed in the V-band  at its BP  (Figure \ref{fig19}, bottom) show tails and streams of a distorted/warped  galaxy.

\section{Discussion}

\begin{figure}
  \centering
 {\includegraphics[width=10cm]{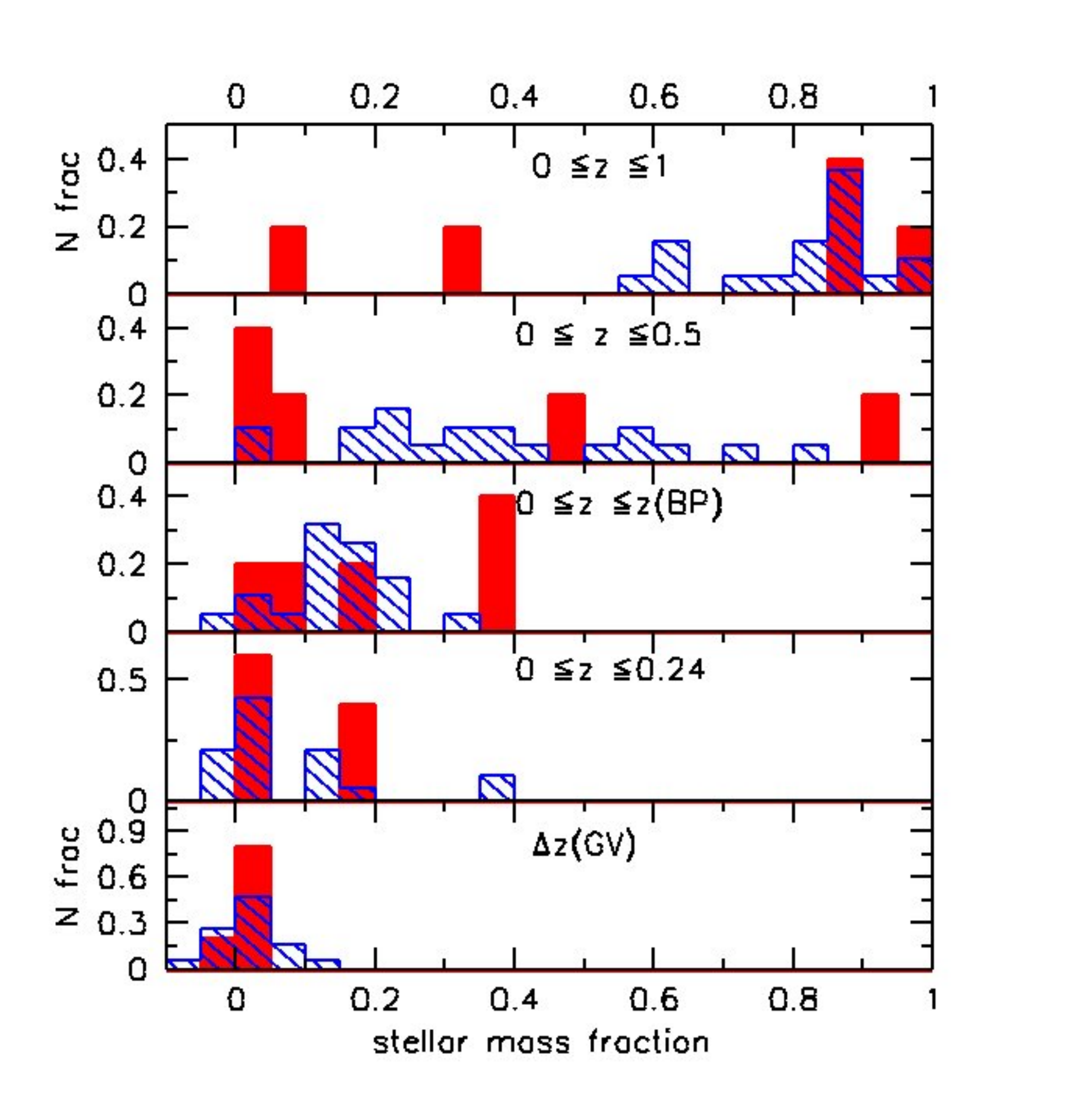}}
\caption{ Fraction of galaxies vs. fraction of stellar mass assembled in different evolutionary steps  within a sphere of radius 50\,kpc  on  B-band luminous centre; redshift intervals are the same as in Table \ref{tabnew}, from col.5 to col.9. Accounting for the cosmological parameters in Section \ref{intro}, the look-back times to redshift 0.5 and 1 are about 5.2 and 8\,Gyr, respectively; BP and GV indicate the Brighter Point  and the Green Valley of the rest-frame CMD of each galaxy. NB: negative mass fractions indicate mass loss in the interval considered.
Red is for CIG galaxies here,  blue hatched regions for 19 ETGs  in Paper II and  Paper I.
}
   \label{fignew}
   \end{figure}
We deepen our study of the galaxy evolution in low density environments
(LDE)  \citep[ ][and references therein]{Mazzei2014a, Mazzei2019} focusing
here on five CIG galaxies.  We are exploring the merger$/$interaction
scenario using a grid of SPH$-$CPI simulations  starting from triaxial halos initially composed of DM and gas \citep{CM98, MC03}  anchored {\it at
posteriori} at the global galaxy properties of our targets to give insight into their
evolution. We already applied this method to study the evolution  to
several galaxy in groups (Section \ref{Sims}).
Mergers and$/$or interactions (encounters) modify the SFR compared to  single collapsing halos  \citep[ ][and references therein]{MC03}  since these mechanisms deepen modify the potential well where the gas is accreting.
The growth of the stellar mass develops from the SF driven by this accretion history.
The SFR  changes for every change of the initial conditions in our grid of simulations. This means that  the evolution of the selected  simulations is quite different  from those in the neighborhood ones in terms of parameter space. The large number of local constrains used allow us to fix the choice.

Are iETGs different from ETGs in LDE? The question is raised by the
discovery that about 60\% of iETGs show shells, indicative of
merging events  \citep[][and references therein]{Rampazzo2020a,Rampazzo2020b}.  \citet{Rampazzo2020a,Rampazzo2020b} also found a variety of
fine structures in the CIG members, not just shells.

Figure \ref{fignew} (top panel) and Table \ref{tabnew} show that the fraction of  stellar mass accreted from redshift 1 by  CIG~189 and CIG~389,  is less than others 3 CIG,   11 ETGs in \citet{Mazzei2019},  and 8 ETGs in \citet{Mazzei2014a}, in agreement with their isolation. However,   looking at the same Figure,  the fraction of stellar mass accreted by CIG here analyzed  in several interesting evolutionary {\bf ranges} is similar to ETGs in LDE \citep[][their Table 4]{Mazzei2019}.

\begin{figure}
  \centering
 {\includegraphics[width=8.8cm]{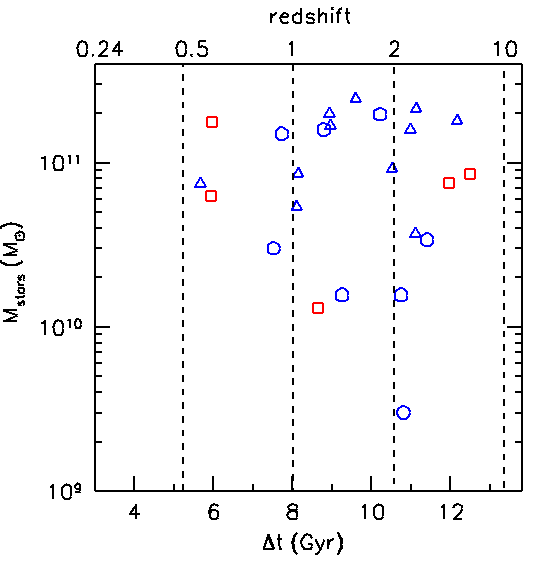}}
\caption{Total stellar mass at $z=0$ vs merger age, that is the time range from the beginnig of the merger (as defined in Section \ref{simul}) to the current galaxy age we derived (Table \ref{table7}).
Five red open squares are results here, eleven blue triangles are from Paper II and eight blue circles from Paper I.
}
   \label{fig16}
   \end{figure}

 Figure\ref{fig16}  shows that our CIG   galaxies  span the full range of
merger-ages of  ETGs  in Paper II, here 0.9$-$ the stellar mass develops from the SF 4.2\,Gyr. This
time-range,   considering  their current age in Table \ref{table7}, corresponds to a
look-back time from  5.9 to 12\,Gyr. 
The youngest mergers, 
corresponding to a look-back time of 5.9\,Gyr,  and 6\,Gyr,
are those occurred in CIG~481 and CIG~309, respectively.
Moreover, these five  systems  span
a large stellar and
total mass range,  7.5$--$17.6$\times$10$^{10}$\,M$_{\odot}$, and 9.5--55.6$\times$10$^{10}$\,M$_{\odot}$ respectively. CIG~637 and
CIG~481 are the youngest galaxies based on their current ages in Table
\ref{table7}. 

\begin{figure*}
  \centering
{\includegraphics[width=8.8cm]{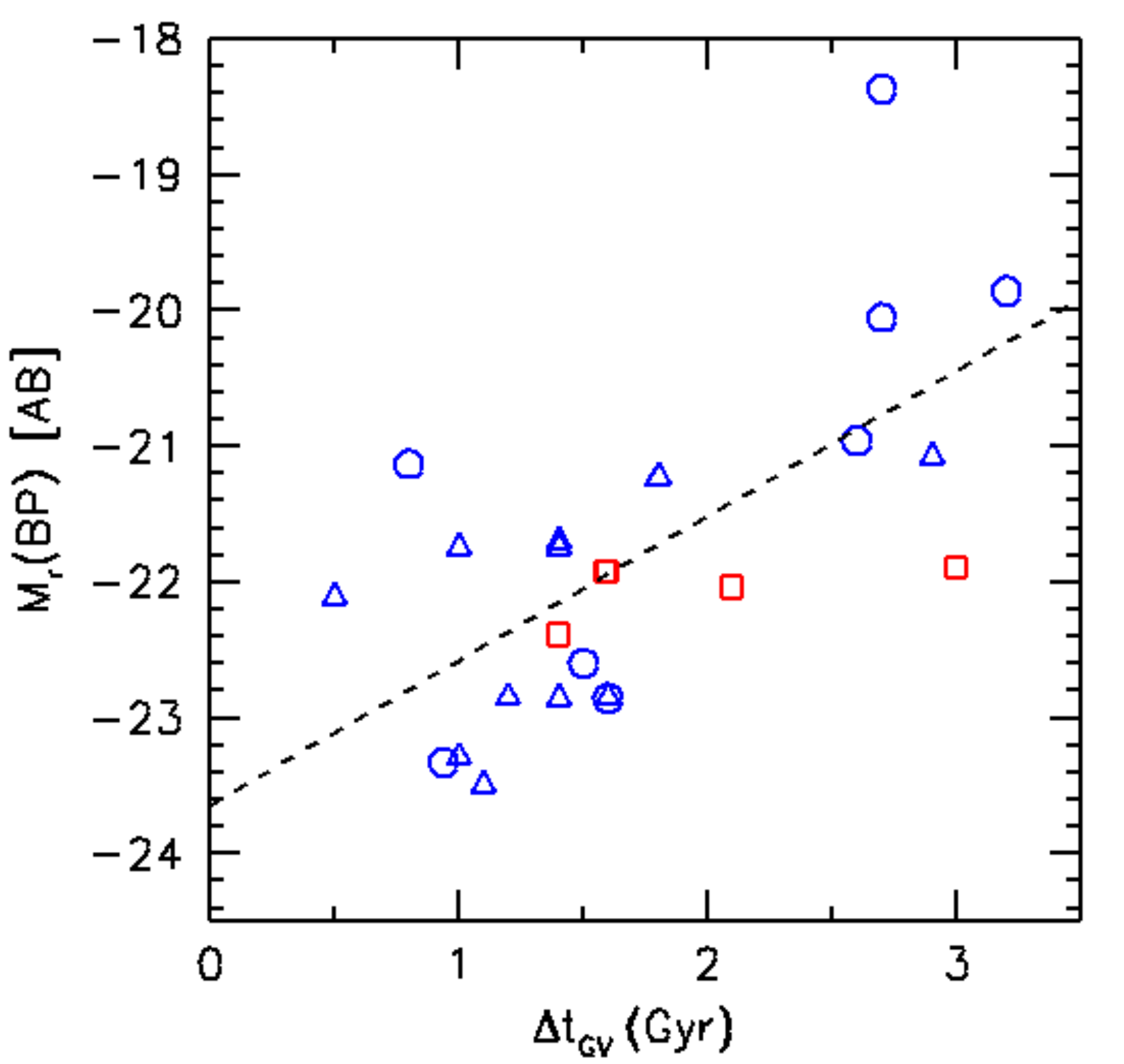}}
 {\includegraphics[width=8.8cm]{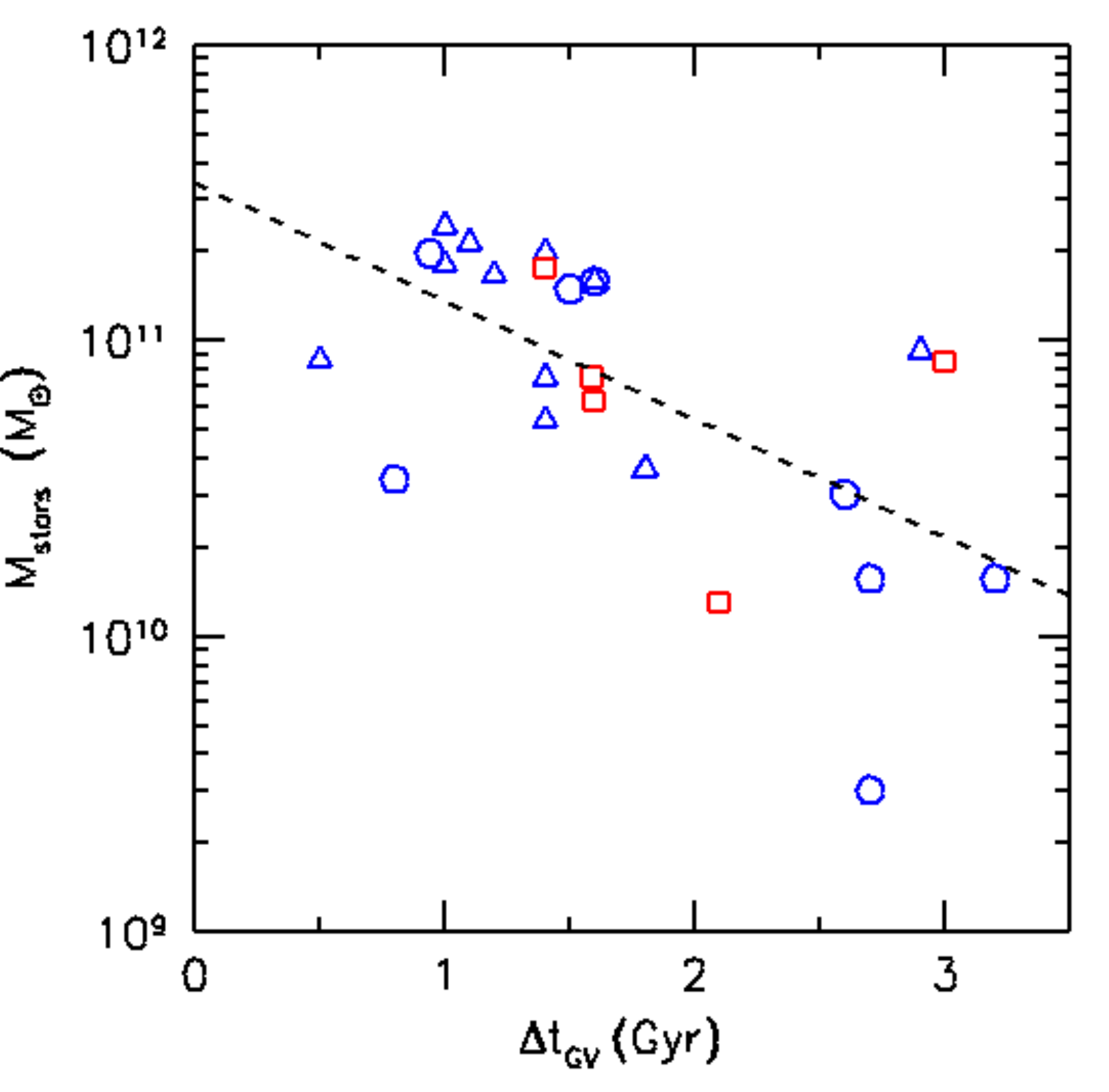}}
\caption{(Left) Absolute M$_r$ [AB] magnitude reached at the BP of the rest-frame CMD vs. time spent in the GV, as defined in Section \ref{simul}. The correlation index is 0.66, and the regression (dashed line) is  $M_r(BP)=-23.65+1.07\times\Delta t_{GV}$.
(Right) Total stellar mass at $z =0$  (Table \ref{table7}) vs. time spent in the GV. The correlation index is -0.61 with  $M_{stars}=11.53-0.4 \times\Delta t_{GV}$. Symbols are as in Figure \ref{fig16}.
}
 \label{fig17}
 \end{figure*}

Their crossing time of the GV anti-correlates with their  local stellar mass
with a good anti-correlation index, (-0.61)  (Figure
\ref{fig17}, right),  and the magnitude at the BP correlates well with the
same time range (Figure \ref{fig17}, left). The evolution derived by our simulations of five CIG
galaxies   is driven by the stellar mass trough SFR and related feedback.
The merger occurred in the past of the galaxy, produced the potential
well and gas reservoir that drive the gas assembly history. The growth
of the stellar mass develops from the SFR driven by this accretion
history. The quenching, which gives rise to galaxy transformation,
occurs from the behavior of the SFR, that is, gas exhaustion and stellar
feedback, several gigayears after the start of the merger. Therefore the
 SF quenching is not due to  the merger, rather,  it is a consequence
of the galaxy evolution. Our
results suggest that the AGN does not affect the global evolution of our
targets. This does not mean that these galaxies have no AGNs.
Their feedback effects cloud be important in the nuclear
regions, at spatial resolution below that considered here (50\,pc), 
in agreement with the
pictures of \citet{Bremer18} and \citet[][see also  \citet{Mazzei2019} for a discussion]{Eales2017, Eales2018a,
Eales2018b}.
The SN feedback is enough to allow quenching of all our
targets.  Their   stellar masses range from 3.7 to 24.4
$\times$10$^{10}$\,M$_{\odot}$ while their total masses range from 9.5 to
61.2$\times$10$^{10}$\,M$_{\odot}$ accounting for the results here and
those by  \citet{Mazzei2019}. Therefore, the evolution expected for CIG galaxies
proceeds through  the same steps as for  galaxies in LDE.

We note
 that, despite   the good isolation validation of CIG~481 (Section
 \ref{isol}), its observed projected morphologies at z=0.24  are very
 different from those expected for a normal ETG, at odds with CIG~189
 which shows   morphologies in agreement with the idea of  an isolated
 ETG  (Figure \ref{fig18}). The morphology of CIG~ 309 is quite peculiar, showing a  plume at
 the same redshift, and those of CIG~389 and CIG~637 are similar to
 barred galaxies. 
We note that  all the CIG galaxies here show complex
morphologies at the BP of their rest-frame CMD (Figure \ref{fig19}), 
with tails and features like jelly-fish galaxies \citep{Roman2019}. This point will be explored further
in a future paper.

Our simulations suggest that 1) the mechanism that drives the evolution 
of galaxies in LDE is the merging and 2) that there is  a
substantial continuity in the evolution of grouped and isolated ETGs. This  is shown 
by Figures \ref{fig16}  and \ref{fig17}, this latter  summarizing  evolution in terms of CMD.    We can 
therefore argue that the current isolation of our targets is the final byproduct of the
evolution in a galaxy-poor environment,  starting from a pair or  triplet halos.
In this framework we may attempt to explain a set of observations. {\it i)} The high 
percentage of shells (60\%) found in iETGs, compared to richer environments 
\citep{MC83,Reduzzi1996, Rampazzo2020a} suggests that, since the last merging episode, 
no further 'external' merger episodes occurred.  If not disturbed, shells are 
therefore long-lasting fine structures  as suggested by observations of \citet[][and references therein]{Longhetti98a,Longhetti98b}
and confirmed by recent simulations of \citet{Mancillas2019} which derive a lifetime of 4\,Gyr for these features.

A further possible consequence of isolation is that the large fraction of rings   revealed
in the set of iETGs investigated  by \citet{Rampazzo2020a} 
should be produced by  resonances, as we already found for NGC~1533 \citep[][ and references therein]{Mazzei2019} and CIG~309 in this paper.
 {\it ii)}  The presence of mergers at redshift  larger than 0.5, highlighted in Figure \ref{fig16}, 
and the evolutionary path we derived,  like those of galaxies in group, indicate 
that isolated galaxy samples are not useful to infer differences with 
interacting galaxies \citep{Lisenfeld2011}.  
The galaxy transformation to ETGs, faster in more massive galaxies, 
it is driven by a quenching process which is, however, a gentle process, lasting more than 1 Gyr.
Therefore ETGs and star forming galaxies belong to the same galaxy population observed
at different evolutionary stages.

\section{Summary and Conclusions}

{\tt SWIFT-UVOT} observations in six bands, from 0.15
$\mu$m  to  0.55\,$\mu$m, of five  CIG galaxies are presented.  
From the
analysis of their  brightness luminosity profiles we derive new S\'ersic's
indices and AB integrated magnitudes (Table \ref{table3} and Table \ref{table6} respectively). In particular, 
we obtain new
UV  morphologies and luminosity profiles  in three bands, from 0.15  to 0.3\,$\mu$m, of all our targets. 
New U
magnitudes for CIG~389 and CIG~637, and V-band magnitude for CIG~189 are
also derived. 
The S\'ersic's indices between 1 and 2 in all these bands for CIG~309 and CIG~389 show a dominant disk morphology.
CIG~189, CIG~481, and CIG~637 have S\'ersic's indices greater than 2.5, as for ETGs  (Section \ref{resobs}.) 
All these new data, together with available literature
information (Appendix), are used to constrain, {\it a posteriori}, our SPH-CPI
simulations to give  insight into their evolution. From our grid of simulations of mergers and encounters,
starting from triaxial collapsing systems of DM and gas, we select those
matching the current global properties of each galaxy, in particular i) the
absolute B-band magnitude, ii) the SED  extended over four orders of magnitude in wavelength, and iii) the optical and UV morphologies as
confirmed by iv) their corresponding luminosity profiles. Moreover, each selected
simulation  accounts for v) the HI gas mass, and vi) the available kinematical data of the target
galaxy.  
No constrains on the amount of hot gas are derived by the literature for our targets.

These simulations correspond to  a major merger in 3 cases and 
a minor merger \citep[mass ratio $>$ $4:1$ as in ][]{Mazzei2014a,
Mazzei2019}  for  CIG~189 and CIG~389. All  mergers occurred 
before $z=0.5$, that is more than 5.2\,Gyr  in the galaxy past (Figure
\ref{fig16}), in agreement with their definition of isolated systems (Section  \ref{isol}).
 CIG~189,  and CIG~389, are the oldest galaxies
(13.7\,Gyr, Table \ref{table7})  in our sample. Therefore they are expected to be passively evolving.
However the merger  is still
leaving its signatures on the evolution of all CIG galaxies here, as shown by 
several rejuvenation episodes in their rest-frame CMD,  triggered, of course, by {\it in situ} SF.

The evolution of these five isolated galaxies  is not different from
evolution of 19 galaxies in LDE we analyzed \citep{Mazzei2019, Mazzei2014a}. This is mass dependent and  does not
require any non-thermal feedback. Our simulations, which include stellar
feedback from SNae and stellar winds, show  that   {\it in situ} evolution,
driven by the gas accretion history, self-regulates the SFR that pushes
the galaxy evolution.  External (e.g., ram pressure) or
additional (e.g., AGN) quenching are not required. Brighter
and more massive galaxies spend a shorter time in the GV than fainter,
less massive ones. Both these conclusions are based on the results of
the SPH-CPI simulations presented here, anchored to the current
properties of 5 CIGs  and of 19 ETGs in nearby groups 
whose stellar mass range,
0.3-24.4$\times$10$^{10}$\,M$_{\odot}$, is about two orders of magnitude.
Moreover, we find that complex morphologies appear as the galaxy achieves  its BP of the rest-frame CMD (Figure \ref{fig19}), 
showing  tails and features like jelly-fish galaxies \citep{Roman2019}. 
Both isolated  and  interacting galaxy samples provide  objects  following the same evolutionary path but observed in different stages of their evolution and
galaxy transformation.
IETGs and ETGs in group are not unperturbed galaxies.  
Their evolution is driven by the evolution of their potential well that can still be evolving, depending on the initial conditions of the merger.

\section*{Acknowledgments}
We thank the referee for helpful comments improving our manuscript.
We acknowledge the use of public data from the {\it Swift} data archive.
We acknowledge use of the {\tt HyperLeda} data base (http://galaxies1.univ-
lyon1.fr). This research has made use of the NASA/IPAC Extragalactic
Database (NED), which is operated by the Jet Propulsion Laboratory,
California Institute of Technology, under contract with the National
Aeronautics and Space Administration. 
We acknowledge the usage of the AKARI data
(http://www.ir.isas.jaxa.jp/AKARI/Publications/guideline.html). The
facility used is {\it Swift}.\\

\begin{appendix}
\section{Additional information from the literature}
\label{add}

Kinematical and gas richness information about our galaxies 
are extremely useful to identify the galaxy model 
from  the grid of the available SPH-CPI simulations.
We summarize below those we found in the literature and used in the  sections above. 

For CIG~189 there are no 
information other than those included in Section \ref{sim189}.

\bigskip
\noindent {\it CIG~309.} \citet{Fabricius2012} derived an optical velocity
dispersion of 173.9$\pm$13.7\,km\,s$^{-1}$  within the inner 1\farcs7.
As a comparison, the {\tt Hyperleda} catalog gives
$\sigma_c$=171.2$\pm$3.8\,km\,s$^{-1}$ and a maximum gas rotational
velocity corrected by inclination (40$^\circ$) of 296.1$\pm$8.5\,km\,
s$^{-1}$. H$\alpha$ imaging from \citet[][their Figure D31]{Epinat2008} 
 comparable with that from \citet{Hamed2005}, shows a flocculent ring of
emission with a maximum rotation velocity of 337$\pm$20\,km\,s$^{-1}$.
However, \citet{Kregel2005} suggest a maximum velocity of 283\,km,
s$^{-1}$ lower than that  of \citet{Epinat2008} (see their Fig. E5), and
like  the value of  {\tt Hyperleda} above.

 We point out that estimating the  distance is  very complex for this galaxy,
 given its position in the Local velocity field (Section \ref{isol}). As
 an example, \citet{Epinat2008} assume a distance of 17.1\,Mpc and a
 B-band absolute magnitude of -20.3\,mag, whereas \citet{Fabricius2012}
 give 14.4\,Mpc and a B-band absolute magnitude of -19.8\,mag, almost 1 mag fainter than the value in Table \ref{table1}.

 \bigskip
 \noindent {\it CIG~389.}   Following \citet{Cappellari2011} the distance of this galaxy is 23.0\,Mpc. On this basis
\citet{Serra2012} derived un upper limit to the cold gas mass,
log(M$_{HI}$[M$_{\odot}$])$<$7.12. rescaling to the distance in
Table~\ref{table1} we derive log(M$_{HI}$[M$_{\odot}$])$<$7.19
(Table~\ref{table1}). Following \citet[][their Table 1]{Molaeinezhad17},
the maximum central velocity dispersion is 126.1\,km\,s$^{-1}$,  in
agreement with the value in the {\tt Hyperleda} catalog,  112.6$\pm$5.1\,
km\,s$^{-1}$. The mean SSP equivalent age  is 14.0$\pm$0.61\,Gyr (their
Table 2). The star and gas maximum rotation velocities are
140.0$\pm$15.8 and  129.7$\pm$22.0\,km\,s$^{-1}$ respectively from {\tt
Hyperleda} catalog.

\bigskip
\noindent{\it CIG~481.}  \citet{Courtois2015} measure 
F$_{HI}$=2.32 \,J\,km\,s$^{-1}$ for this galaxy  which, rescaling to the distance in Table 1,  gives
log(M$_{HI}$[M$_\odot$])=8.49. The maximum rotational velocity of  the
gas, corrected for  inclination (56.8$^\circ$), is
152.4$\pm$5.7\,km\,s$^{-1}$ by {\tt HyperLeda} catalog.

\bigskip
\noindent {\it CIG~637.} Following \citet{Cappellari2011} its distance is 27.2\,Mpc. They measure
a mass of cold gas, log(M$_{HI}$(M${_\odot}$)=7.32 that, rescaling to
the distance in Table~\ref{table1},  gives
log(M$_{HI}$[M$_{\odot}$])=7.50. Moreover, \citet{Costantin2018}
conclude  this is a classical bulge galaxy with a luminosity-weighted
value of the line-of-sight velocity dispersion  within an elliptical
aperture equal to the effective radius (6.67\arcsec, their Table 4) of
170$\pm$6\, km\,s$^{-1}$.

\end{appendix}

\end{document}